\documentclass[twocolumn]{aastex62}
\newcommand{\tblspc}{-0.1cm}
\usepackage[normalem]{ulem}
\usepackage[version=4]{mhchem}

\received{...}
\revised{...}
\accepted{...}
\shorttitle{Icy pebbles feeding inner disk chemistry}
\shortauthors{Banzatti et al.}

\begin{document}

\title{Hints for icy pebble migration feeding an oxygen-rich chemistry in the inner planet-forming region of disks}

\correspondingauthor{Andrea Banzatti}
\email{banzatti@txstate.edu}

\author{Andrea Banzatti}
\affil{Department of Physics, Texas State University, 749 N Comanche Street, San Marcos, TX 78666, USA}
\affil{Department of Planetary Sciences, University of Arizona, 1629 East University Boulevard, Tucson, AZ 85721, USA}

\author{Ilaria Pascucci}
\affil{Department of Planetary Sciences, University of Arizona, 1629 East University Boulevard, Tucson, AZ 85721, USA}
\affil{Earths in Other Solar Systems Team, NASA Nexus for Exoplanet System Science}

\author{Arthur D. Bosman}
\affil{Department of Astronomy, University of Michigan, 1085 S. University Ave, Ann Arbor, MI 48109}

\author{Paola Pinilla}
\affil{Max-Planck-Institut f\"ur Astronomie, K\"onigstuh l17, 69117 Heidelberg, Germany}

\author{Colette Salyk}
\affil{Department of Physics and Astronomy, Vassar College, 124 Raymond Avenue, Poughkeepsie, NY 12604, USA}

\author{Greg J. Herczeg}
\affil{Kavli Institute for Astronomy and Astrophysics, Peking University, Beijing 100871, China}

\author{Klaus M. Pontoppidan}
\affiliation{Space Telescope Science Institute 3700 San Martin Drive Baltimore, MD 21218, USA}

\author{Ivan Vazquez}
\affil{Department of Physics, Texas State University, 749 N Comanche Street, San Marcos, TX 78666, USA}
\author{Andrew Watkins}
\affil{Department of Physics, Texas State University, 749 N Comanche Street, San Marcos, TX 78666, USA}

\author{Sebastiaan Krijt}
\affil{Department of Astronomy, The University of Arizona, 933 North Cherry Avenue, Tucson, AZ 85721, USA}
\affil{Earths in Other Solar Systems Team, NASA Nexus for Exoplanet System Science}
\affiliation{Hubble Fellow}

\author{Nathan Hendler}
\affil{Department of Planetary Sciences, University of Arizona, 1629 East University Boulevard, Tucson, AZ 85721, USA}

\author{Feng Long}
\affiliation{Harvard-Smithsonian Center for Astrophysics, 60 Garden Street, Cambridge, MA 02138, USA}


\begin{abstract}
We present a synergic study of protoplanetary disks to investigate links between inner disk gas molecules and the large-scale migration of solid pebbles. The sample includes 63 disks where two types of measurements are available: \textit{i)} spatially-resolved disk images revealing the radial distribution of disk pebbles (mm-cm dust grains), from millimeter observations with ALMA or the SMA, and \textit{ii)} infrared molecular emission spectra as observed with \textit{Spitzer}. The line flux ratios of \ce{H2O} with \ce{HCN}, \ce{C2H2}, and \ce{CO2} all anti-correlate with the dust disk radius R$_{\rm{dust}}$, expanding previous results found by \citet{naj13} for \ce{HCN}/\ce{H2O} and the dust disk mass. By normalization with the dependence on accretion luminosity common to all molecules, only the \ce{H2O} luminosity maintains a detectable anti-correlation with disk radius, suggesting that the strongest underlying relation is between \ce{H2O} and R$_{\rm{dust}}$. If R$_{\rm{dust}}$ is set by large-scale pebble drift, and if molecular luminosities trace the elemental budgets of inner disk warm gas, these results can be naturally explained with scenarios where the inner disk chemistry is fed by sublimation of oxygen-rich icy pebbles migrating inward from the outer disk. Anti-correlations are also detected between all molecular luminosities and the infrared index n$_{13-30}$, which is sensitive to the presence and size of an inner disk dust cavity. Overall, these relations suggest a physical interconnection between dust and gas evolution both locally and across disk scales. We discuss fundamental predictions to test this interpretation and study the interplay between pebble drift, inner disk depletion, and the chemistry of planet-forming material.
\end{abstract}

\keywords{circumstellar matter --- protoplanetary disks --- stars: pre-main sequence --- }

\section{INTRODUCTION} \label{sec:intro}
The Atacama Large Millimeter/Submillimeter Array (ALMA) has revolutionized our understanding of the outer regions of protoplanetary disks \citep[beyond tens of au; see e.g.][for a review]{andr20}. Pronounced dust substructures demonstrate that disks are highly diverse and dynamical systems, and may suggest that planet formation is well underway in the Class II stage \citep[e.g.][]{dsharp2,long18}. Overall, disk images observed at millimeter wavelengths, that probe the presence and radial distribution of disk pebbles (mm/cm-size dust grains), point to dust growth to pebbles and their inward radial drift as key ingredients in disk evolution and planet formation \citep[e.g.][]{testi14,pasc16,pin12,pin20}. In the emerging pebble accretion formation scenario, \citet{lamb19} suggest that it is the inward flux of migrating pebbles that determines whether a planetary system will form the numerous super-Earths detected by \textit{Kepler} or rather smaller planets like Earth.
Icy pebbles migrating inward from the outer disk are also expected to alter the volatile content of the inner rocky planet-forming zone within 10~au \citep[e.g.][]{cc06,krijt18,krijt20,bosm18,booth19}.
Therefore, both observations and theoretical predictions point toward potential strong interconnections between disk evolution and planet formation processes across disk scales. Investigating these interconnections requires the combination of observatories that trace different disk regions (inner versus outer disk).
While ALMA is best suited to spatially resolve substructures at tens to hundreds of au, the inner region within 10~au is instead best probed via infrared (IR) observations of atomic and molecular spectra \citep[e.g.][for a review]{pont14}. 

In this work, we study correlations between mid-infrared molecular spectra as tracers of inner disk chemistry and spatially-resolved measurements of dust disk radii as a tracer of the radial distribution of solid pebbles. Infrared molecular spectra have revealed a forest of emission lines from CO, \ce{H2O}, OH, as well unresolved ro-vibrational bands from HCN, \ce{C2H2}, and \ce{CO2} observed in protoplanetary disks, especially around young T~Tauri stars \citep[e.g.][]{cn11,salyk11a,salyk11b,pont10a,mand12,naj13, pasc13,brow13,banz17}. While most studies have focused on the analysis of inner disk gas tracers and stellar or inner disk properties, \citet{naj13} reported a positive correlation between the HCN/\ce{H2O} flux ratio from \textit{Spitzer} spectra and the dust disk mass as estimated from millimeter observations. This finding is particularly remarkable because it links the disk mass tracing dust grains in the outer disk ($> 20$~au) and molecular spectra tracing the gas within a few~au from the star. The authors proposed it as evidence for locking of \ce{H2O} ice into large planetesimals and planetary cores beyond the snow line, increasing the C/O ratio in the inner disk region. They suggested that this would happen more efficiently in more massive disks, as they have more solid mass to form planetesimals and planets that accrete and lock water ice beyond the snow line. With this interpretation, the authors proposed that inner disk molecules might provide a ``chemical fingerprint" of planetesimal formation that is happening in the outer disk \citep{naj18}.

This work is motivated by the findings reported in \citet{naj13} and by the dramatic improvement in resolution and quality of millimeter disk images that happened since. In this work, we aim at expanding the analysis of \citet{naj13} by: 1) including a $\sim 3$ times larger disk sample (from 22, counting only those that had millimeter disk mass measurements, to 63 disks), 2) studying correlations for four molecules instead of two (\ce{H2O}, HCN, \ce{C2H2}, and \ce{CO2}), and 3) studying correlations with spatially-resolved millimeter observations of  dust disk radii, instead of disk masses estimated from integrated millimeter fluxes and SED fits. 
Recent surveys have shown that the total millimeter flux and the outer disk radius correlate well \citep{trip17,long19,hend20}, although the origin of this relation is still unclear \citep{andr20}. When the millimeter flux is converted into an estimate of disk mass, usually assuming a fixed factor given by a constant dust opacity and an average dust temperature\footnote{\citet{andr13} and \citet{pasc16} also explored a disk temperature dependence on stellar luminosity, but this dependence is still uncertain and is at most weak \citep{tazz17}.}, a correlation between disk radius and mass is also found \citep{trip17}. However, the derivation of disk masses from millimeter fluxes is now more than ever highly debated, due to uncertainties in the opacities and optical depth of the dust and to dust trapping in substructures that make a simple derivation unreliable \citep[e.g.][]{ric12,birn18,dull18,andr20}. Therefore, in this work we focus on spatially-resolved measurements of the outer dust disk radius rather than on disk mass estimates, because radii provide a more direct measurement of a fundamental disk property, i.e. the radial extent of pebbles, without the dependence on the several assumptions that go into estimating disk masses\footnote{For instance, in the millimeter dust masses from SED fits by \citet{aw05}, adopted by \citet{naj13}, the estimated dust mass depends also on the uncertain disk temperature structure.}.

This paper is structured as follows. In Section \ref{sec: data} we present the sample properties and the data. Millimeter disk radii are adopted from recent surveys (Sect. \ref{sec: radii}), and infrared molecular line fluxes are measured from spectra reduced in previous work (Sect.\ref{sec: spectra}). Section \ref{sec: results} presents the analysis of correlations between molecular line luminosities and stellar and disk properties. In Section \ref{sec: discuss} we discuss these results in the context of the drift efficiency of icy pebbles from the outer disk feeding an oxygen-rich inner disk chemistry, and the formation of inner disk cavities. We conclude with predictions for future work, focusing on how the results and interpretation from this analysis can be further tested and expanded with future synergies of high-resolution data.

\begin{figure*}
\centering
\includegraphics[width=1\textwidth]{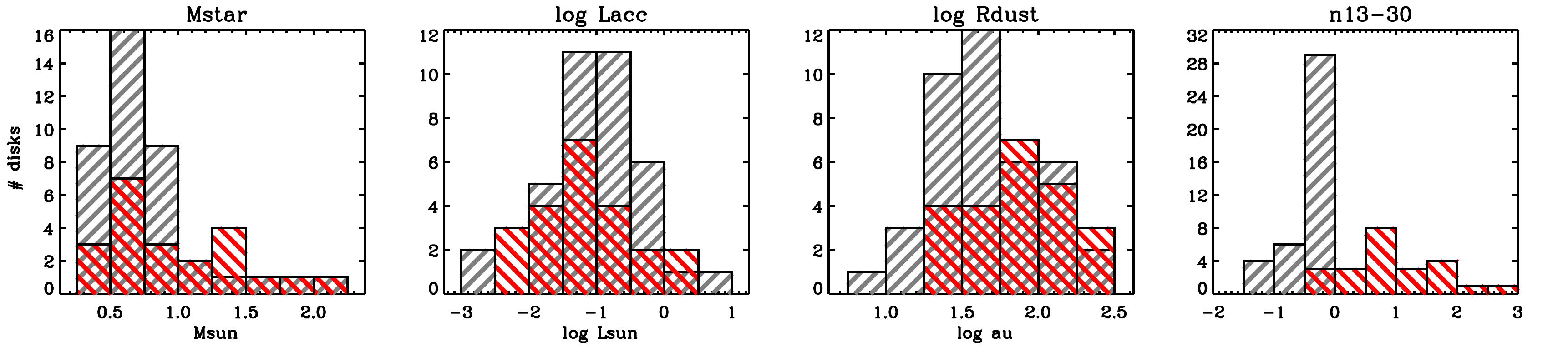} 
\caption{Sample property distributions (see Section \ref{sec: data} for details). Full disks are shown in grey; disks with inner cavities are shown in red. }
\label{fig: sample}
\end{figure*}

\section{Sample \& Data} \label{sec: data}
\subsection{Sample selection and properties}
The sample analyzed in this work includes 63 protoplanetary disks around pre-main-sequence stars (see Appendix \ref{app: sample}) that currently have two types of measurements available: 1) dust disk radii measurements from recent high-resolution surveys using ALMA or the SMA (see details in Section \ref{sec: radii}), and 2) high-quality $Spitzer$ spectra covering mid-infrared molecular gas emission (Section \ref{sec: spectra}). The sample includes T\,Tauri stars from nearby ($< 200$ pc) star-forming regions of similar age (1--3 Myr): Taurus, Lupus, Ophiuchus, Chamaeleon I. Ten disks are in known binary or multiple stellar systems (these are discussed in Appendix \ref{app: binaries}). We exclude from this work disks around Herbig A/B stars, because these are known to have predominantly no mid-infrared molecular lines detected \citep{pont10a} possibly due to dissociation processes related to the stronger irradiation field and to the presence of large cavities \citep[e.g.][]{fed11,banz18,bosm19}. 

In terms of molecular spectra, the sample includes the full range from the strongest measured line emission (typically from gas- and dust-rich disks around $\sim 0.5-1$ M$_{\odot}$ stars) down to upper limits from disks that have inner dust cavities \citep{pont10a,salyk11a}. In terms of disk dust radii, the sample includes the full range that has been spatially resolved with ALMA or the SMA, from $> 200$\,au down to $\sim 10$\,au \citep[see][for a review]{andr20}. The sample also includes 24 disks that have an inner dust cavity; 16 of these cavities have been spatially resolved with high-resolution millimeter imaging (see Table \ref{tab: sample}), other cavities are inferred from the infrared index $n_{13-30}$ \citep{brow07,furl09}. In Appendix \ref{app: IRindex}, we discuss $n_{13-30} > 0$ as tracing inner disk dust cavities, and highlight its dependence on the disk inclination that in close to edge-on orientations can lower $n_{13-30}$ to less than 0 even in the presence of an inner dust cavity. The spectral index $n_{13-30}$ is measured as in \citet{banz19} from narrow spectral ranges that avoid molecular emission at 13.1~$\mu$m and 30.1~$\mu$m. Accretion luminosity measurements are taken from \citet{fang18} and \citet{sim16} for roughly half the sample (30 disks), and from \citet{salyk13} and other works for the rest of the sample. Sample properties and all references are included in Table \ref{tab: sample}, and the distributions of sample properties are shown in Figure \ref{fig: sample}.

\subsection{Dust disk radii}  \label{sec: radii}
Measurements of disk dust radii are adopted from recent intermediate- to high-resolution surveys from millimeter interferometers. These surveys obtained spatially-resolved disk images at 1.3 mm with ALMA with angular beams of $\sim 0.''04-0.''12$ \citep[][available for $\sim75\%$ of the sample studied in this work]{dsharp2,long19}, or at 0.9 mm with ALMA or the Submillimeter Array (SMA) with angular beams of $\sim 0.''3-0.''7$ \citep[][available for $\sim25\%$ of the sample]{trip17,andr18a,hend20}. The analysis of disk images has been performed in the interferometric visibility space, providing sub-beam resolution and spatially-resolved disk size measurements for the entire sample included here. In this work we use their measurements of the radius R$_{\rm{dust}}$ that encircles 90\% or 95\% (depending on what reported in the original papers) of the observed millimeter continuum flux. \citet{trip17} and \citet{andr18a} reported the radius that includes the 68\% of the flux, instead, and we use the relation reported by \citet{hend20} to derive the radius that encircles 90\% of the flux, to be consistent with the rest of the sample. \citet{hend20} demonstrated the applicability of this relation for the range of angular resolutions used in this work; the 1-$\sigma$ uncertainty on disk radii from this relation is $\lesssim 0.1$~dex. Disk radii are included in Table \ref{tab: sample}.

\begin{figure*}
\centering
\includegraphics[width=0.9\textwidth]{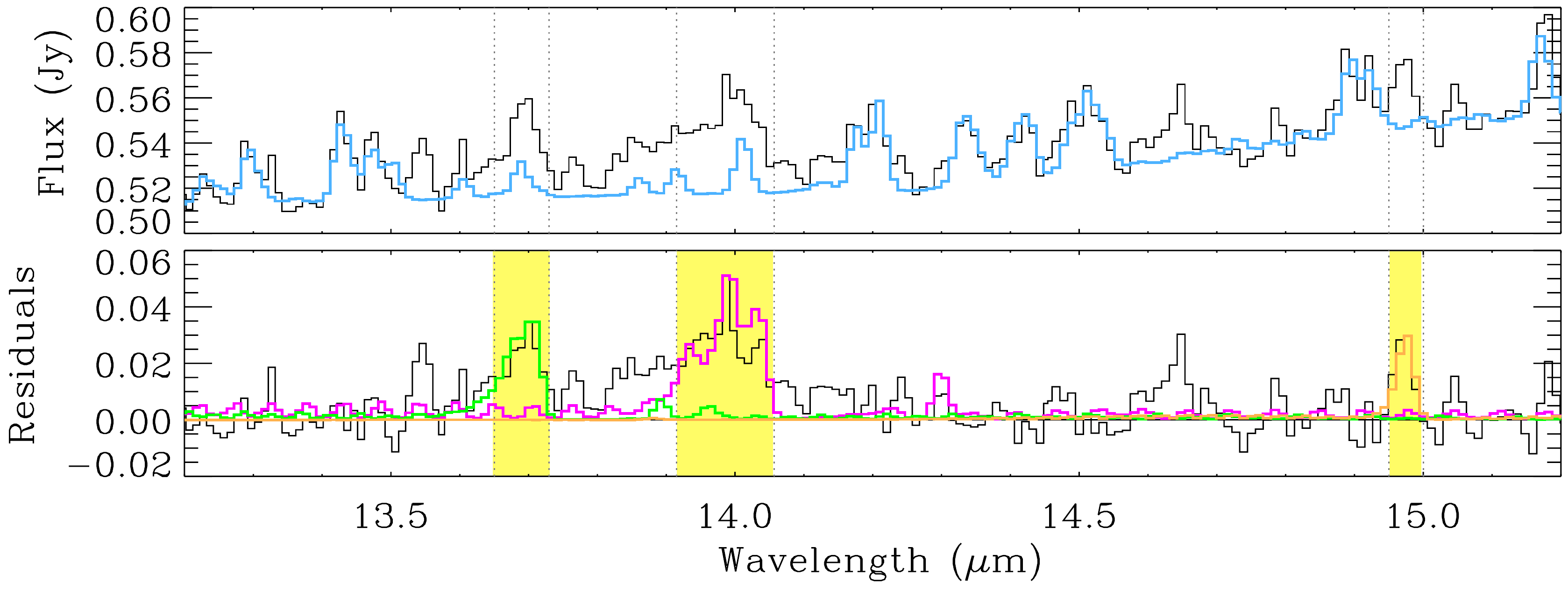} 
\caption{Correction of carbon-bearing molecular features from water contamination in $Spitzer$-IRS spectra. The spectrum of CI\,Tau is shown as an example. \textit{Top}: observed spectrum in black, water emission model in cyan (see Section \ref{sec: spectra}). \textit{Bottom}: residuals after subtraction of the water emission model from the data. Models of carbon-bearing molecular emission features are shown for guidance (but are not fits to the data): \ce{C2H2} at 13.7\,$\mu$m in green, HCN at 14\,$\mu$m in magenta, \ce{CO2} at 14.95\,$\mu$m in orange. The yellow shaded area indicates the ranges where carbon-bearing molecular line fluxes are measured.}
\label{fig: spitzer_line_fluxes}
\end{figure*}

\begin{figure}
\includegraphics[width=0.47\textwidth]{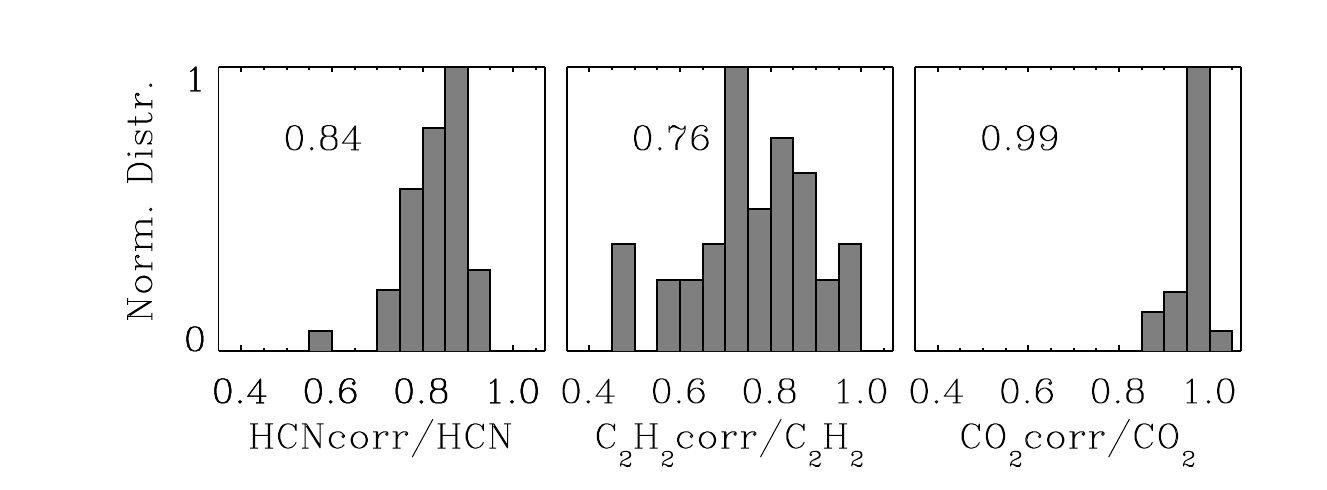} 
\caption{Distributions of the ratio of corrected line flux to original (uncorrected) line flux. The median value of each distribution is reported to the top left.}
\label{fig: orgcorr_histos}
\end{figure}

\subsection{Gas emission line fluxes} \label{sec: spectra}
The molecular spectra analyzed in this work were taken with the \textit{Spitzer}-IRS high-resolution modules \citep{IRS} and we adopt the reduced data from \citet{pont10a} and \citet{rigl15}. Additional spectra are taken from the CASSIS database \citep{cassis} for disks with available measurements of millimeter dust radii: Elias\,24 and CV\,Cha with water and/or carbon-bearing molecules detected, and five disks that have no molecular detections or only \ce{CO2}: CS\,Cha, MY\,Lup, Sz\,73, SR\,4, UX\,Tau\,A. Details on data reduction procedures can be found in the original references listed above. 

Studying molecular spectra taken with the \textit{Spitzer}-IRS presents challenges that have been extensively discussed in previous work. These challenges are mostly due to the low spectral resolution of the \textit{Spitzer}-IRS SH and LH modules \citep[$R\sim720$ as measured from unresolved hydrogen lines,][]{naj10,banz13} that blends multiple emission lines together and produces a pseudo-continuum of weak emission lines \citep[e.g.][]{cn11,liu19}, the low pixel sampling (typically only a few pixels per each emission line blend, in the case of water; see e.g. Figure \ref{fig: spitzer_line_fluxes}), and residuals from fringe removal \citep[see e.g. discussion in][]{pont10a}. Due to these factors, some emission line detections are marginal or only tentative, especially in spectra with weak molecular emission. We adopt best practices developed in previous works on these spectra \citep[e.g.][]{pont10a,cn11,banz12}, where the local continuum is determined from pixels that have no or the weakest emission lines (as determined using molecular emission models) and consider lines detected only if above 3$\sigma$. In Section \ref{sec: future}, we will discuss how these issues will be solved or at least mitigated with future higher-resolution data. 
 
As in \citet{naj13}, the water line fluxes used for the analysis are taken as the sum of three well-separated emission features at 17.12~$\mu$m, 17.22~$\mu$m, and 17.36~$\mu$m. The flux in these features is dominated by transitions with very similar upper level energy and Einstein coefficient \citep[2400--3300 K and 1--4 s$^{-1}$ respectively; see e.g. Table 3 in][with data taken from the HITRAN database, \citet{hitran}]{banz17}, such that they meet very similar excitation conditions. These lines are typically observed to have very similar peak-to-continuum contrast \citep[e.g. Figure 6 in][]{pont10a}, supporting the expectation that they probe a very similar portion of the emitting gas.
Line fluxes for the carbon-bearing molecules are measured over the ranges showed in Figure \ref{fig: spitzer_line_fluxes}. HCN line flux measurements include the strongest part of the branch around 14~$\mu$m, and avoid water contamination in the shorter-wavelength tail of the feature, similarly to the procedure adopted in \citet{naj13}. 

To measure line fluxes of the carbon-bearing molecules, we first remove water emission (where present) from the spectra with the following procedure. We take a model of a slab of gas in LTE \citep[described in][]{banz12}, where the water emission spectrum is defined by two parameters: the excitation temperature T and column density N (the emitting area is just a scaling factor for the whole spectrum). We take T = 600\,K and N = $10^{18}$ cm$^{-2}$, which \citet{cn11} found to fit well the water spectrum observed at 12--16~$\mu$m in several T~Tauri disks. We measure the local continuum with two linear fits anchored over the following wavelength ranges: 13.38--13.41~$\mu$m, 14.26--14.28~$\mu$m, and 15.00--15.03~$\mu$m, and then apply these fits to the slab model to match the continuum flux and slope at these wavelengths (Figure \ref{fig: spitzer_line_fluxes}).
We then scale the peak-to-continuum strength of the model to match the observed water emission around the HCN line (using \ce{H2O} lines at 13.43~$\mu$m and 14.19--14.35~$\mu$m) in each spectrum, in order to account for different emitting areas in different disks \citep[see e.g.][]{salyk11a}. The carbon-bearing molecular line fluxes are measured from the residuals after subtraction of the continuum + water emission model, over the ranges visualized in Figure \ref{fig: spitzer_line_fluxes} \citep[where we include, just for guidance, models of carbon-bearing molecular emission using average T and N as reported by][]{cn11}. Figure \ref{fig: orgcorr_histos} shows the distribution of fractional corrections (the ratio between corrected and uncorrected line flux), demonstrating that \ce{CO2} is the least affected and \ce{C2H2} the most affected by water contamination, as it can be seen also from the example in Figure \ref{fig: spitzer_line_fluxes}. All line flux measurements are reported in Table \ref{tab: fluxes}. In the next Section, we use molecular line luminosities $L$ in units of solar luminosity, defined from the measured line fluxes $F$ as $L = 4 \pi d^2 F$, where $d$ is the distance (Table \ref{tab: sample}).

\begin{figure*}
\centering
\includegraphics[width=0.75\textwidth]{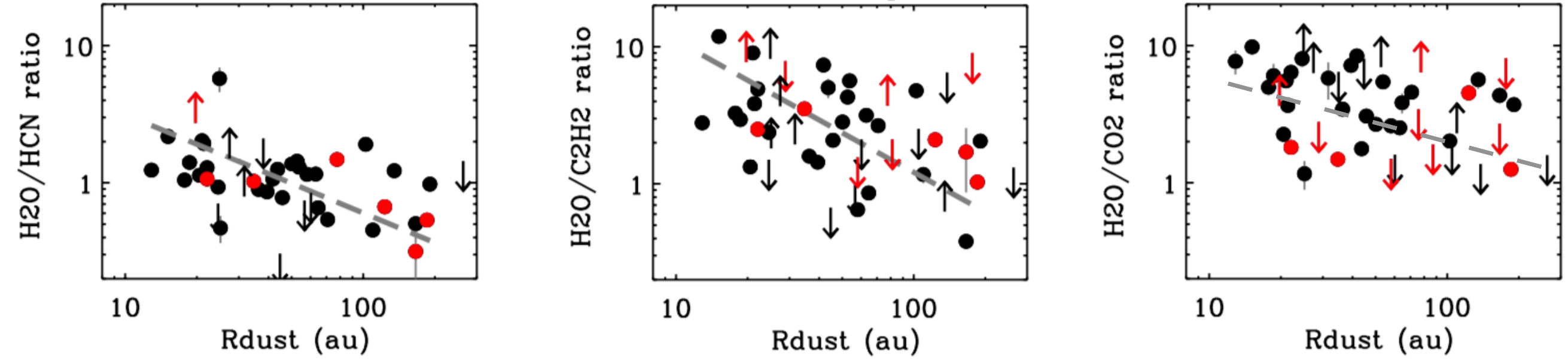}
\caption{Correlations between water/carbon-bearing-molecules line flux ratios and millimeter disk radii (see Section \ref{sec: data} for details). Red datapoints identify disks that have an inner dust cavity. Linear regression fits are shown with dashed lines.}
\label{fig: corr_molratios}
\end{figure*}

\begin{figure*}
\centering
\includegraphics[width=0.95\textwidth]{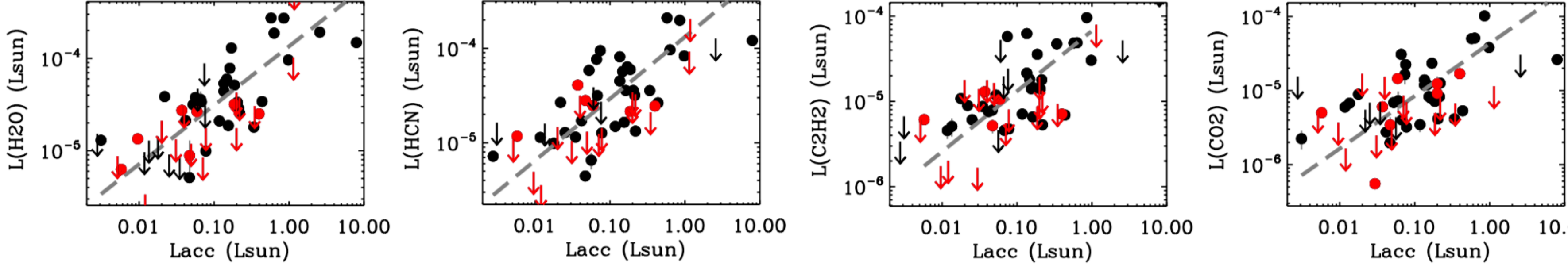}
\caption{Correlations between molecular line luminosity and accretion luminosity. Linear regression fits are shown with dashed lines. All fits are consistent with a power law of index 0.6 (Table \ref{tab: correlations}), as $L_{\rm{molecule}} \propto L_{\rm{acc}}^{0.6}$.}
\label{fig: corr_Lacc}
\end{figure*}

\begin{figure*}
\centering
\includegraphics[width=0.95\textwidth]{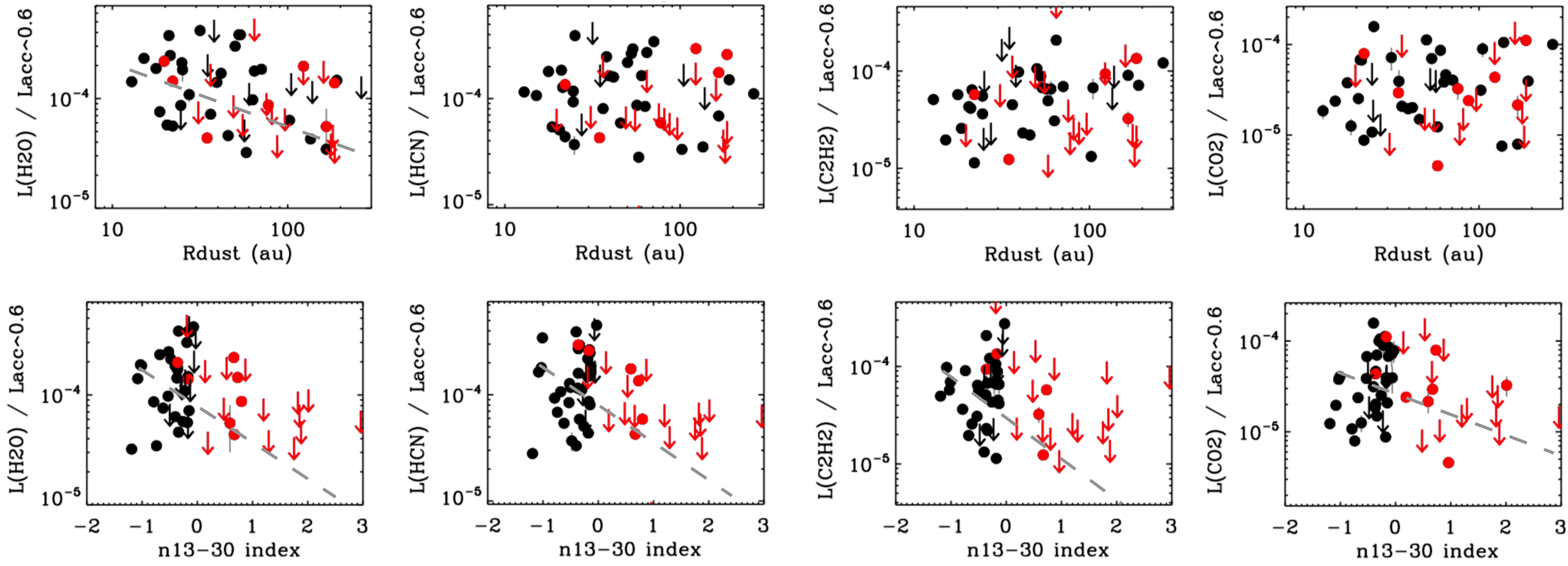}
\caption{Correlations between molecular luminosity divided by $L_{\rm{acc}}^{0.6}$ and disk radius (top) or infrared index (bottom). Linear regression fits are shown with dashed lines where a correlation is detected in the data.}
\label{fig: corr_Laccratios}
\end{figure*}

\begin{deluxetable*}{l c c c c c c c c c}
\tabletypesize{\footnotesize}
\tablewidth{0pt}
\tablecaption{\label{tab: correlations} Linear regression results.}
\tablehead{\colhead{ } & \multicolumn{3}{c}{log L$_{\rm{acc}}$ (L$_\odot$)} & \multicolumn{3}{c}{log R$_{\rm{dust}}$ (au)} & \multicolumn{3}{c}{n$_{13-30}$}
\\ 
 & $\alpha$ & $\beta$ & $\hat \rho ~|~ \sigma_{\epsilon}$ & $\alpha$ & $\beta$ & $\hat \rho ~|~ \sigma_{\epsilon}$  & $\alpha$ & $\beta$ & $\hat \rho ~|~ \sigma_{\epsilon}$}
\tablecolumns{10}
\startdata
\ce{H2O} & \textbf{-3.97$\pm$0.12} & \textbf{0.60$\pm$0.10} & \textbf{0.73 $|$ 0.4} & -3.73$\pm$0.47 & -0.62$\pm$0.27 & -0.35 $|$ 0.6 & 
\textbf{-4.85$\pm$0.10} & \textbf{-0.55$\pm$0.15} & \textbf{-0.63 $|$ 0.6} \\
\ce{HCN} & \textbf{-3.91$\pm$0.11} & \textbf{0.64$\pm$0.09} & \textbf{0.69 $|$ 0.4} & -4.37$\pm$0.40 & -0.23$\pm$0.24 & -0.15 $|$ 0.6 & 
\textbf{-4.83$\pm$0.08} & \textbf{-0.55$\pm$0.13} & \textbf{-0.70 $|$ 0.5} \\
\ce{C2H2} & \textbf{-4.22$\pm$0.11} & \textbf{0.62$\pm$0.09} & \textbf{0.71 $|$ 0.4} & -5.03$\pm$0.38 & -0.06$\pm$0.22 & -0.05 $|$ 0.5 & 
\textbf{-5.19$\pm$0.07} & \textbf{-0.54$\pm$0.12} & \textbf{-0.75 $|$ 0.4} \\
\ce{CO2} & \textbf{-4.55$\pm$0.11} & \textbf{0.65$\pm$0.10} & \textbf{0.62 $|$ 0.4} & -4.88$\pm$0.36 & -0.21$\pm$0.21 & -0.15 $|$ 0.5 & 
\textbf{-5.24$\pm$0.08} & \textbf{-0.28$\pm$0.10} & \textbf{-0.43 $|$ 0.5} \\
\ce{H2O}/HCN & 0.13$\pm$0.14 & 0.14$\pm$0.14 & 0.39 $|$ 0.2 & \textbf{0.54$\pm$0.20} & \textbf{-0.33$\pm$0.12} & \textbf{-0.49 $|$ 0.2} & 
-0.02$\pm$0.06 & -0.01$\pm$0.10 & -0.02 $|$ 0.3 \\
\ce{H2O}/\ce{C2H2} & 0.51$\pm$0.24 & 0.20$\pm$0.22 & 0.37 $|$ 0.4 & \textbf{1.20$\pm$0.30} & \textbf{-0.52$\pm$0.18} & \textbf{-0.50 $|$ 0.3}
 & 0.30$\pm$0.08 & -0.02$\pm$0.14 & -0.04 $|$ 0.4 \\
\ce{H2O}/\ce{CO2} & 0.63$\pm$0.20 & 0.16$\pm$0.17 & 0.36 $|$ 0.3 & \textbf{1.02$\pm$0.25} & \textbf{-0.33$\pm$0.15} & \textbf{-0.40 $|$ 0.3}
 & \textbf{0.36$\pm$0.06} & \textbf{-0.37$\pm$0.11} & \textbf{-0.66 $|$ 0.3} \\
\ce{H2O}/L$^{0.6}_{\rm{acc}}$ & -- & -- & -- & \textbf{-3.24$\pm$0.30} & \textbf{-0.50$\pm$0.18} & \textbf{-0.45 $|$ 0.4} & 
\textbf{-4.11$\pm$0.07} & \textbf{-0.34$\pm$0.10} & \textbf{-0.59 $|$ 0.4} \\
HCN/L$^{0.6}_{\rm{acc}}$ & -- & -- & -- & -3.86$\pm$0.30 & -0.15$\pm$0.17 & -0.14 $|$ 0.4 & \textbf{-4.12$\pm$0.06} & \textbf{-0.35$\pm$0.10}
 & \textbf{-0.63 $|$ 0.4} \\
\ce{C2H2}/L$^{0.6}_{\rm{acc}}$ & -- & -- & -- & -4.65$\pm$0.28 & 0.12$\pm$0.16 & 0.13 $|$ 0.4 & \textbf{-4.49$\pm$0.06} & 
\textbf{-0.36$\pm$0.10} & \textbf{-0.69 $|$ 0.3} \\
\ce{CO2}/L$^{0.6}_{\rm{acc}}$ & -- & -- & -- & -4.43$\pm$0.32 & -0.10$\pm$0.18 & -0.08 $|$ 0.4 & \textbf{-4.59$\pm$0.07} & 
\textbf{-0.21$\pm$0.09} & \textbf{-0.39 $|$ 0.4} \\
\enddata

\tablecomments{
Results from the linear regression test by \citet{kelly07}. The linear relation is in the form $y = \alpha + \beta x + \epsilon$, where $\alpha$ and $\beta$ are the intercept and slope, $\epsilon$ the intrinsic scatter with standard deviation $\sigma_{\epsilon}$, and $\hat \rho$ the correlation coefficient. The dependent variables $y$ are given in the first column and correspond to what shown in Figures \ref{fig: corr_molratios} to \ref{fig: corr_Laccratios}, while the independent variables $x$ are given at the very top in the other columns. For $\alpha$ and $\beta$ we report the median and standard deviation of the posterior distributions. Correlations that are considered detected and significant are marked in boldface. For a comparison to other correlation tests, see Table \ref{tab: correlations_appendix}.
}
\end{deluxetable*}

\section{Correlation analysis and results} \label{sec: results}
Investigating processes that affect the measured molecular line luminosities is intrinsically a multi-dimensional problem. Mid-infrared molecular spectra are expected to strongly depend on gas heating and cooling and their dependence on the inner disk irradiation, geometry, dust/gas density and their vertical/radial distributions, among other factors \citep[e.g.][]{naj11,du14,walsh15,bosm17,woitke16,woitke18}. 
While determining the relative importance of these effects requires detailed modeling, previous observational work found clear evidence for two major effects that overall control the mid-infrared molecular emission. First, molecular luminosities correlate with the stellar luminosity \citep{salyk11a} and accretion luminosity \citep{banz17}, supporting a strong role for gas heating processes. Second, molecular luminosities decrease when inner disk dust cavities form \citep{naj10,salyk15,banz17}, suggesting that inner disk molecular gas gets depleted when the dust is depleted. 
In this work we therefore focus on these two known trends and we use the large sample for a systematic correlation analysis aimed at investigating other underlying effects, especially those related to the radial extent and migration of disk pebbles. 

For comparison to common procedures adopted in other works, correlations are assessed with both the Pearson's correlation test for linear relations and Spearman's rank correlation test for monotonic relations, and we report in the Appendix (Table \ref{tab: correlations_appendix}) their correlation coefficients and the associated two-sided probability of the data not being correlated (p-values). We adopt the common cut of a probability smaller than 5\% to consider a correlation detected in the data. However, as in \citet{hend20} we remark that both these tests have their limitations in capturing correlations, one being the fact that they cannot account for measurement uncertainties and upper limits. Therefore, to better assess correlations we adopt the Bayesian method for linear regression by \citet{kelly07} using the \texttt{linmix\_err} code, which does account for upper limits and uncertainties on both the dependent and independent variables. This method has been shown to reproduce the results of other common statistical tests that include censored data \citep{kelly07,pasc16}. This method is particularly suited for multi-dimensional problems, as it accounts for an intrinsic scatter in the linear relation due to physical properties that are not explicitly included in the variables (e.g. when the measured molecular luminosity is affected by multiple factors). The linear relation is in the form $y = \alpha + \beta x + \epsilon$, where $\alpha$ and $\beta$ are the intercept and slope and $\epsilon$ the intrinsic scatter with standard deviation $\sigma_{\epsilon}$. We use this method for the linear regressions included in Table \ref{tab: correlations}. This method does not estimate p-values but provides posterior distributions for the model parameters and for the correlation coefficient ($\hat \rho$), from which we measure median values to represent the best fit results.
In this work we adopt a lower limit of 0.4 in the absolute value of the correlation coefficient to consider a correlation detected in the data (Table \ref{tab: correlations}); just for guidance, this value would correspond to a p-value of $\approx$~2--3\% in the Pearson's correlation test. 
\\

\subsection{Water/carbon-bearing molecule flux ratios} \label{sec: mol_ratios}
In reference to \citet{naj13}, we first report the correlations between water/carbon-bearing molecule flux ratios and disk radii in Figure \ref{fig: corr_molratios}.
An anti-correlation is detected between the flux ratio \ce{H2O}/HCN and R$_{\rm{dust}}$, supporting the correlation found earlier by \citet{naj13} with disk mass. In addition, this analysis detects a similar anti-correlation between \ce{H2O}/\ce{C2H2} and R$_{\rm{dust}}$, and similar also for \ce{H2O}/\ce{CO2}, although scatters are larger (Table \ref{tab: correlations}). 

We remark that these correlations are robust regardless of whether the carbon-bearing molecular features are corrected for water contamination or not, confirming what was reported by \citet{naj13} for the HCN/\ce{H2O} ratio. They are also robust against different choices of the pixels used to measure the local continuum, whether the ranges used here or those used in \citet{naj13}. 

We note that we do not detect correlations between line luminosities for the individual molecules and R$_{\rm{dust}}$, in this sample (but see Section \ref{sec: lum_corr} for the molecular luminosities corrected for the accretion luminosity). Similarly, \citet{naj13} reported no correlation between \ce{H2O} or HCN line fluxes and disk dust masses, and suggested that the correlation is in the relative molecular abundance rather than in the individual molecules. Yet, the interpretation of these line flux ratios is still unclear, as they may reflect changes in excitation conditions, optical depth, and emitting disk regions that can be different for different molecules. To aid the interpretation of these correlations, we therefore analyze the underlying correlations between molecular luminosity and stellar/accretion luminosity.

\subsection{Luminosity normalization of IR molecular spectra} \label{sec: lum_corr}
In a previous analysis of \textit{Spitzer} spectra, \citet{salyk11a} found that mid-IR molecular luminosities are correlated with stellar luminosity, and explained these correlations as an emitting area effect where the radial extent of the observed emission varies from disk to disk depending on the irradiation from the star. Using IR emission lines observed over a larger wavelength range (2.9--33 $\mu$m) and considering multiple molecules (\ce{H2O}, OH, and CO), \citet{banz17} also found correlations between molecular luminosities and the accretion luminosity or the total (stellar + accretion) luminosity.  

Here, we repeat the correlation analysis and determine that the strongest correlation is with the accretion luminosity  (Figure \ref{fig: corr_Lacc}). Correlations with $L_{\star}$ consistently present coefficients lower by $\sim$~50\% (from $\sim$~0.7 to $\sim$~0.3) and a larger intrinsic scatter by $\sim$~20--50\% (from $\sim$~0.4 to $\sim$~0.5--0.6 dex), and the case is similar with $L_{\rm{tot}}$ (coefficients lower by $\sim$~40\% and scatter larger by $\sim$~20--40\%). The stronger correlation with $L_{\rm{acc}}$ could be due to the fact that the gas is directly heated by UV photons that dominate the accretion spectrum \citep[e.g.][]{du14,woitke18}. All molecular luminosities in Figure \ref{fig: corr_Lacc} scale with $L_{\rm{acc}}$ with a similar power law index of $\sim0.6 \pm 0.1$, suggesting that they respond similarly to excitation from the accretion luminosity. For reference, this dependence to $L_{\rm{acc}}$ is weaker than for hydrogen and helium optical lines that are typically used as accretion tracers \citep[e.g.][]{alc17,fang18}, but it is similar to the dependence found in mid-infrared hydrogen lines \citep{rigl15}.

These strong correlations suggest that we should first remove the accretion luminosity dependence before we can investigate other processes that affect the observed molecular lines. To do so, we divide the measured line luminosities by $L_{\rm{acc}}^{0.6}$. We find that this luminosity normalization provides an effective correction in removing any correlation with $L_{\star}$ and $L_{\rm{tot}}$ too \citep[this is not surprising because $L_{\rm{acc}}$ and $L_{\star}$ are correlated, e.g.][]{alc17}. 
After removing this luminosity effect, we investigate what else controls the observed molecular emission.

\subsection{Normalized molecular luminosity and R$_{\rm{dust}}$}
Once normalized to the accretion luminosity, only \ce{H2O} still shows a significant anti-correlation with R$_{\rm{dust}}$ (Figure \ref{fig: corr_Laccratios}, top), with a coefficient of $-0.45$ consistent with those reported above for the molecular ratios and R$_{\rm{dust}}$ (Table \ref{tab: correlations}). We also note that among the four molecules, \ce{H2O} presents the highest correlation coefficient (an absolute value of 0.35) even before normalization with $L_{\rm{acc}}$, further supporting the existence of a relation between \ce{H2O} and R$_{\rm{dust}}$. It is also worth noting that, currently, the upper limits measured in some disks play an important role in driving this correlation, as they populate the trend at the large disk radii end, where disks with inner cavities (and no molecular emission) tend to be found (see also Figure \ref{fig: sample}). 

The different behavior of the four molecules, where only water shows a detectable correlation, suggests that \ce{H2O} might have the strongest relation with R$_{\rm{dust}}$. The scatter is large compared to the slope of the relation, but the Bayesian regression measures a similar scatter of $\sim0.4$ dex in all four molecules, once normalized by $L_{\rm{acc}}$ (Table \ref{tab: correlations}). This suggests that if a similar correlation were present in all of them it would be detected, but does not exclude the presence of a shallower correlation in the carbon-bearing molecules. To further test the detection of a correlation with \ce{H2O}, we run a bootstrapping procedure to estimate a false-positive rate. We run this test in two ways, in one case by randomly shuffling the measured y-values and in another case by randomly drawing y-values modeled as having the same intrinsic scatter as measured in the data, but no relation with the x-values. In both cases, we keep the measured x-values and run the bootstrap simulation 10,000 times, re-fitting the data in each realization with the Bayesian regression method used above. These tests provide us with two distributions of 10,000 realizations of the $L_{\rm{molecule}}$-R$_{\rm{dust}}$ relations in the assumption that there is no relation between the two variables. In both cases, we measure a false-positive rate of $\sim1$\% of finding an absolute value of the correlation coefficient equal or larger than 0.4, suggesting high confidence on the \ce{H2O}-R$_{\rm{dust}}$ anti-correlation detected in the data.

Therefore, the \ce{H2O}-R$_{\rm{dust}}$ anti-correlation detected in the data suggests that the main driver of all correlations found in the water/carbon-bearing molecular ratios (Section \ref{sec: mol_ratios}) may be \ce{H2O}, without excluding a weaker dependence for the other molecules that is worth investigating in future work (as in fact supported by models, see Section \ref{sec: disc_drift}).

\subsection{Normalized molecular luminosity and $n_{13-30}$} \label{sec: n1330_corr}
A persistent anti-correlation, this time in common to all four molecules, is found between the accretion-normalized line luminosity and the infrared index $n_{13-30}$ (Figure \ref{fig: corr_Laccratios}, bottom). Differently from the case with R$_{\rm{dust}}$, these anti-correlations are detected even before normalizing line luminosities by the accretion luminosity (Table \ref{tab: correlations}). 
The dependence of infrared line detections to $n_{13-30}$ in T~Tauri disks has been known since the analysis in \citet{salyk11a}, where a correlation was found with detection rates for \ce{H2O}, \ce{OH}, and \ce{CO2}, and marginally also for \ce{HCN} and \ce{C2H2}.
Here, by including the upper limits, we expand the previous analysis from line detections to line luminosities, and confirm the anti-correlation in all molecules. By considering separately disks with or without an inner cavity, we find that this anti-correlation remains for all molecules in disks with cavities (with the caveat that the sample is composed of upper limits and only 5--7 detections). No correlation is found in the full disks if considered by themselves. It is therefore possible that these correlations appear only in disks with inner cavities, but we should also note that $n_{13-30}$ values span a much larger range in these disks as compared to those without a cavity, making it easier to detect any correlations with this specific infrared index if they exist. It is also interesting to note that, given their partial overlap in terms of molecular luminosity and $n_{13-30}$, full disks could still represent the initial conditions of cavity disks.

These correlations suggest that $n_{13-30}$ traces processes that have a strong effect on the inner disk molecular gas as a whole. These results add up to a series of new correlations with $n_{13-30}$ found in other inner disk gas tracers, specifically in [OI] emission at 6300~\AA \citep{banz19} and [NeII] emission at 12.81~$\mu$m \citep{pasc20}. In comparison, while the correlations with [OI] emission are all dominated by detections \citep{banz19}, the mid-infrared molecular tracers still largely rely on upper limits currently measured in disks with an inner dust cavity ($n_{13-30} > 0$), identifying a clear venue for future improvement (see Section \ref{sec: future}).

\begin{figure*}
\centering
\includegraphics[width=0.95\textwidth]{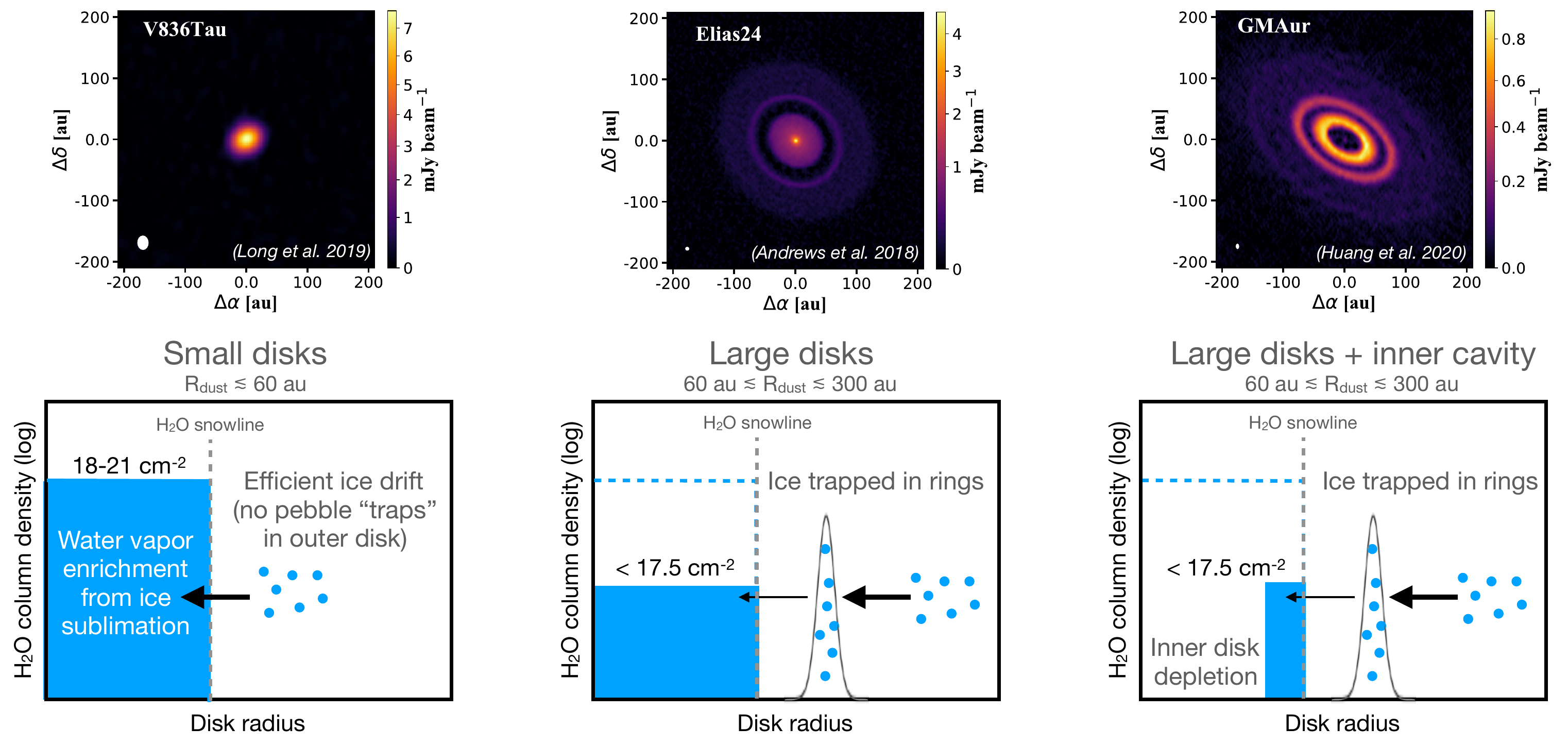} 
\caption{Cartoon to visualize our interpretation of the results from this analysis. Efficient drift of icy pebbles in small disks leads to large columns of water vapor inside the snow line, and therefore to a high infrared \ce{H2O} luminosity (Section \ref{sec: results}). Large disks have substructures where icy pebbles are trapped in the outer disk, lowering the columns of inner water vapor and decreasing the \ce{H2O} luminosity. Similar or lower columns are found in disks with an inner cavity, where dust and gas are depleted. Indicative column density values for the three size-bins of disk radii are taken from fits to water emission in \textit{Spitzer} spectra \citep[][and Appendix \ref{app: water_coldens}]{salyk11a}. The images to the top show representative disks from \citet{long19} for small disks, \citet{dsharp1} for large disks with substructures, and \citet{huang20} for large disks with a cavity.}
\label{fig: cartoon}
\end{figure*}

\section{Discussion} \label{sec: discuss}
In this work, we have analyzed correlations between mid-infrared molecular line luminosities, their ratios, and fundamental star and disk properties. In reference to previous results by \citet{naj13}, we have analyzed correlations between water/carbon-bearing molecular flux ratios and spatially-resolved measurements of disk dust radii (rather than disk mass), finding an anti-correlation between \ce{H2O}/HCN ratio and R$_{\rm{dust}}$. The analysis also reveals that similar anti-correlations, although with larger scatter and possibly different slopes, exist for \ce{H2O}/\ce{C2H2} and \ce{H2O}/\ce{CO2}. After normalization to remove the accretion luminosity dependence common to all molecular lines, only an anti-correlation between \ce{H2O} luminosity and R$_{\rm{dust}}$ remains, suggesting that carbon-bearing molecular fluxes mostly acted as normalization factors and that the main driver of all correlations is the inner disk \ce{H2O} and its link to R$_{\rm{dust}}$. Moreover, the analysis finds anti-correlations between the luminosity of all four molecules and the infrared index $n_{13-30}$, correlations that are independent of the accretion luminosity normalization applied above.

In this section, we discuss these results in the context of the enrichment or depletion of molecules in the inner disk as linked to pebble drift from the outer disk and the formation of inner disk dust cavities. We illustrate our overall interpretation in Figure \ref{fig: cartoon}, where we include representative ALMA images for the range of disk radii and structures observed in this sample. The interpretation we discuss here relies on two key parts, \textit{i)} that current R$_{\rm{dust}}$ measurements reflect a disk radius that is primarily set by pebble drift at the age of these systems (1--3~Myr), and \textit{ii)} that the range of measured molecular luminosities (once corrected for the measured accretion luminosity) reflects a range of column densities and/or elemental mass budgets in inner disks. Both these components are currently being studied and discussed in the community, and the interpretation for both is still evolving and is the focus of ongoing and future work. Here we report the main arguments that have been discussed in previous work, and how they may provide a natural explanation for the correlations reported in Section \ref{sec: results}. In this section, we focus the discussion on \ce{H2O} which shows the strongest relation with R$_{\rm{dust}}$, but as we said above the current scatter in the data does not exclude shallower relations with the carbon-bearing molecules too. At the end of this section, we also propose key predictions to validate or correct the interpretation illustrated in Figure \ref{fig: cartoon}.
\\

\subsection{Pebble drift, disk substructures, and water enrichment in inner disks} \label{sec: disc_drift}
The correlation between \ce{H2O} and R$_{\rm{dust}}$ is most remarkable, as it links two observables that are completely independent of each other. Firstly, mid-infrared \ce{H2O} line fluxes strictly probe the disk region at $\lesssim 5$~au \citep{pont10b,salyk19}, which well matches the expected location of the water snow line for the stars in this sample \citep[][based on the viscous snow line modeled in \citet{muld15}]{banz17} as well as the snow line location estimated in a few disks from spectral mapping \citep{blev16}. R$_{\rm{dust}}$ values, instead, measure the outer radial extent of pebbles in disks at tens of au, up to 100--200~au. The correlation between \ce{H2O} and R$_{\rm{dust}}$ therefore suggests a link between inner and outer disk regions. Secondly, this correlation shows a connection between dust and gas, in particular between the distribution of pebbles in the disk midplane and the gas content of the inner disk atmosphere. While it cannot be excluded that these correlations may be due to a third unrelated factor that is not currently considered, it is interesting to discuss the possibility of a physical process that may link the observables.

The most natural physical explanation of the \ce{H2O}-R$_{\rm{dust}}$ correlation may be offered by the growing understanding of high-resolution millimeter images of dust emission in disks. It has long been proposed that disks must have some kind of pressure ``traps" in place to prevent pebbles from drifting very rapidly onto the star, well before $\sim1$\,Myr when dust pebbles are observed in disks at $\sim$~10--100~au \citep{pin12}. These pressure traps are now routinely inferred from high-angular-resolution millimeter images of disks that show radial substructures such as rings and gaps \citep[e.g.][]{dsharp1,dsharp2,long18}. These observations have revealed that there is a strong connection between a large outer disk radius and the presence of large-scale substructures in dust emission; in fact, all disks with R$_{dust} > 60$~au, if observed at high angular resolution, have shown dust rings and/or gaps \citep[e.g.][]{dsharp2,long19}. The presence of substructures solves long-standing problems associated to the fast radial drift of solids in disks, and is now dramatically changing our understanding of disk structure and evolution \citep[][for a review]{andr20}. In fact, modeling work proposes that current measurements of R$_{\rm{dust}}$ are mostly set by dust and pebble drift \citep[e.g.][]{ros19a,ros19b}. In this scenario, small R$_{\rm{dust}}$ are indicative of efficient drift that has removed solids from outer disk regions enriching the inner disk. Large R$_{\rm{dust}}$, instead, have pressure variations often associated with planet-disk interactions \citep[e.g.][]{bae18,zhang18,lod19} that maintain a significant amount of pebbles in the outer disk by preventing their migration toward the star \citep[e.g.][]{dull18,pin20}. 

The scenario where the radial drift of pebbles is fundamental in setting the observed disk properties has recently gained increasing support from observations, from the time evolution of millimeter fluxes versus stellar masses \citep[e.g.][]{pasc16,pin20}, to the widespread presence of substructures \citep[][for a review]{andr20}, to the large difference between disk radii as measured in the gas versus the dust \citep{trap19,trap20,facch19,kurt20}. This latter point, in particular, may help to address the important underlying assumption that disks are all born with a similar size, or at least that the relative differences in measured R$_{\rm{dust}}$ are set by dust evolution and migration more than by different initial conditions, which can still contribute to the observed scatter. Measuring disk radii at their formation in Class 0 and I objects is very challenging both for technical reasons and for the presence of dust envelopes \citep[e.g.][]{tobin15,tobin20,zhao20}. 
A key observable to confirm R$_{\rm{dust}}$ as being set by pebble drift in Class II disks is therefore the outer disk gas radius R$_{\rm{gas}}$, which should most closely reflect the disk radius at formation. Although R$_{\rm{gas}}$ measurements are more difficult than R$_{\rm{dust}}$ measurements and are currently available only for a small fraction of the mm-imaged disks, measurements of large R$_{\rm{gas}}$/R$_{\rm{dust}}$ ratios seem to be the most promising venue for ongoing and future studies of pebble drift in disks \citep[][see also Section \ref{sec: future}]{trap19,trap20,facch19,kurt20}.

If R$_{\rm{dust}}$ is in fact a good proxy for the large-scale efficiency of pebble drift in disks, the \ce{H2O}-R$_{\rm{dust}}$ correlation is indicative that water vapor in the inner disk is linked to the efficiency of inward drift of pebbles from the outer disk. Inner disk enrichment by sublimating icy solids that cross the snowline is an effect that has long been proposed in the context of the Solar System \citep{cyr98,cz04}. Early models predicted that the water vapor abundance within the snowline should be tightly linked to the flux of inward migrating icy bodies, which is expected to be strong early on and decrease as disk material is accreted onto the star or onto planetesimals and planets \citep{cc06}. Interestingly, despite how fundamental it is, this prediction has long eluded direct confirmation from disk observations. The present work, by finding a link between infrared water emission and the location of disk pebbles, might be providing the most direct evidence to date for inner disk molecular gas being fed by sublimation of migrating icy solids.

There are two main pathways in which the sublimation of ices can enrich the gas. Either the ice sublimates and directly changes the abundances in the inner disk, or after the ice has sublimated chemical processes destroy and reform molecules into a new kinetic equilibrium. In the former case, it is the molecular composition of the ice that matters, while in the latter it is the elemental composition. In the case that the observations directly probe the sublimating ice composition a strong effect in the \ce{HCN} and \ce{CO2} emission is expected. Analysis of \textit{Spitzer} spectra have found that the inner disk abundances of \ce{HCN} and \ce{CO2} are generally $\sim$0.01\% of the \ce{H2O} abundance \citep{salyk11a, pont14fadi}. The expected abundance of \ce{CO2} and \ce{HCN} in the ice is significantly higher. Cometary observations find HCN/\ce{H2O} ratio of $\sim$1\% and \ce{CO2}/\ce{H2O} of $\sim$10\% \citep{mum11}, and \ce{CO2} is also found at the 10\% level in interstellar ices \citep{boog15}. With these ice abundances, it takes less sublimating ice to significantly change the HCN and \ce{CO2} abundances than to change the \ce{H2O} abundance \citep[see also][]{bosm18}. In this case, if molecular luminosity traces abundance, we would expect that the correlations HCN-R$_{\rm{dust}}$ and \ce{CO2}-R$_{\rm{dust}}$ should be stronger than the \ce{H2O}-R$_{\rm{dust}}$ correlation. As the opposite is observed (Section \ref{sec: results}), it is likely that the observations are not directly probing the sublimating ice.

This leaves chemical processing of the sublimated ices as the most likely pathway. The region that is probed with mid-infrared observations has a high density and is strongly irradiated by the stellar and accretion UV radiation \citep{woitke18}. As such, chemical timescales are short and it is thus likely that the probed gas is in kinetic equilibrium and has lost track of the molecular composition of the enriching gas \citep[see the ``chemical reset" scenario discussed in][]{pont14}. What matters in this case is the change in elemental composition of the gas, both the absolute C/H and O/H ratios as the C/O ratio. Ices feeding the inner disk are expected to be dominated by oxygen carrying molecules, \ce{H2O} and \ce{CO2} \citep[e.g.][]{viss09,bvd12,clee18}. This would, in the case of efficient drift, increase the O/H ratio and lower the C/O ratio \citep[see e.g.][]{bosm18, booth19}. Both \citet{naj11} and \citet{woitke18} predict that the infrared \ce{H2O} luminosity is the most sensitive to changes in the elemental C/O ratio between 0.2 and 0.8, the range expected for the high columns of water observed in inner disks. \ce{HCN} and \ce{C2H2} abundances, instead, are limited by the availability of C and N, which is not strongly impacted by the addition of very O-rich ice. 

Finally, the \ce{CO2} abundance is expected to be nearly linearly dependent on CO and thus the elemental C abundance \citep[e.g.][]{bosm18}. The lack of correlation between the \ce{CO2} luminosity and disk radius thus suggests that influx of elemental C in these inner disks does not strongly depend on the pebble drift from the outer disk, which is consistent with gas-phase CO being the dominant carbon carrier within the CO snowline \citep[][]{bosm17,bosm18, zhang18}.

\subsection{Inner disk dust cavities and the depletion of molecular gas} \label{sec: disc_cav}
Although the $n_{13-30}$ index is affected by several properties of inner disks, including disk flaring and the dust grain size distribution, recent work has revealed that the size of an inner dust cavity plays a major role and likely dominates in producing values of $n_{13-30} > 0$ \citep[see disk model grids in][and Appendix \ref{app: IRindex}]{hond15,woitke16,ball19}. In fact, large positive $n_{13-30}$ are found in disks where large inner dust cavities have recently been spatially-resolved in millimeter emission \citep[e.g.][see also Table \ref{tab: sample}]{brow07,salyk09,huang20}. Therefore, correlations between inner disk gas tracers and $n_{13-30}$ have recently been interpreted as due to gas depletion within inner disk dust cavities \citep{banz19,pasc20}, following what previous work suggested based on a prevalent absence of molecular detections in ``transitional'' disks \citep{naj10,pont10a,salyk11a,salyk15}. 

That molecular gas is depleted in inner disks with dust cavities has been supported by several works using different tracers and modeling techniques. Simple slab model fits found lower \ce{H2O} and CO column densities in disks with inner dust cavities, as compared to ``full" disks \citep[][and Appendix \ref{app: water_coldens}]{salyk11a,salyk11b}. Modeling work by \citet{anton16} found that the infrared water spectrum should respond to the formation and size of an inner cavity with a specific spectral signature, where the depletion of hotter to colder gas by increasing the inner cavity size produces a decrease of higher-excitation lines at shorter wavelengths (3--17~$\mu$m) to lower-excitation lines at longer wavelengths (25--35~$\mu$m), a spectral signature that is observed in the data \citep{banz17}. Thermo-chemical model fits to spectrally-resolved infrared CO emission in Herbig disks estimated the gas column density as a function of inner disk radius, and clarified that when not observed, the CO gas column density must be depleted by at least a few orders of magnitude \citep[][]{brud13,carm17,bosm19}. In some cases, spatially-resolved imaging has also shown depletion of CO gas inside dust cavities \citep{pont08,vdmar16,vdmar18}. 

For guidance, we report here column density fits to water emission in \textit{Spitzer} spectra by \citet{salyk11a}, which are discussed in Appendix \ref{app: water_coldens}. \ce{H2O} column densities in the range $10^{18}$--$10^{21}$\,cm$^{-2}$ were measured in disks that now are shown by millimeter imaging to have small R$_{\rm{dust}}$. These high columns are not matched even by the maximum columns explored by \citet{naj11} that require a C/O ratio as low as 0.2, and could imply further enrichment by a large mass-flux of sublimating icy pebbles. Lower \ce{H2O} column densities of $\lesssim 10^{17.5}$\,cm$^{-2}$ (mostly upper limits) were instead estimated in disks that now are shown to have a large R$_{\rm{dust}}$, and in disks with an inner dust cavity (see Appendix \ref{app: water_coldens}). These previous results, although still tentative because they rely on simple slab models, are suggestive that the correlations between \ce{H2O} and R$_{\rm{dust}}$ or $n_{13-30}$ may in fact be linked to the enrichment or depletion of water vapor in inner disks. The analysis of spectrally-resolved infrared CO lines further suggests a scenario where inner molecular gas depletion not only reduces the gas column density but also shifts the emission to a narrower ring of gas beyond the inner cavity, as illustrated in Figure \ref{fig: cartoon} (see \citet[][]{salyk11b}, Figure 4 in \citet{banz15}, and Appendix \ref{app: water_coldens}).

The correlations between molecular luminosities and $n_{13-30}$, pending further analysis of the currently large fraction of upper limits, might imply a depletion of inner disk gas molecules as a gradual process linked to dust depletion, in terms of the inner dust cavity size or possibly of the degree of dust depletion (and the distribution of dust grain sizes) within the cavity. In Appendix \ref{app: IRindex} we show a comparison of this dataset with existing models published in \citet{ball19}, but a dedicated study of $n_{13-30}$ as a function of inner disk cavity structure and size is yet to be done.
In comparison to the correlation with R$_{\rm{dust}}$, which is strong enough to be detected only in the \ce{H2O} luminosity in the data used in this work, it makes sense that similar correlations with $n_{13-30}$ are shown by all molecules, if these are due to some level of global depletion of the inner disk molecular gas. 

Given this strong effect related to inner disk dust depletion, could the \ce{H2O}-R$_{\rm{dust}}$ relation be entirely due to inner disk depletion in large disks rather than inefficient pebble drift? Or in other words, do all disks with large R$_{\rm{dust}}$ also have an inner dust cavity? Disks with an inner dust cavity do tend to also have a larger radius (Figure \ref{fig: sample}), perhaps because cavity formation is linked to the presence of efficient pebble traps in the outer disk \citep[e.g.][]{pin12,pin18}. However, there is no evidence that \textit{all} large disks also have inner cavities. This question needs to be further addressed with future data (see Section \ref{sec: future}), but the upper panel of Figure \ref{fig: corr_Laccratios} shows that disks with or without a cavity show a similar range of molecular luminosities, possibly implying similar columns of warm molecular gas. However, \textit{Spitzer} spectra have still mostly provided only upper limits for disks with inner cavities, while large disks without a (detected) cavity have been detected in molecular emission. It is therefore possible that future more sensitive data will reveal a dichotomy where large disks have lower \ce{H2O} luminosity than small disks, but large disks with an inner cavity have an even lower \ce{H2O} luminosity than large disks in general. This would further support the scenario where the relative decrease of \ce{H2O} in large disks as compared to smaller disks is not only due to an inner cavity, but generally linked to pebble drift as discussed above. The detailed interplay between dust drift and disk cavity formation on inner disk chemistry is still largely to be explored, and will likely require new high-resolution observations of dust in inner disks from the next generation of infrared observatories (see Section \ref{sec: future}).

\subsection{Implications for planet formation through pebble accretion} \label{sec: planet_formation}
It is interesting to discuss the results of this work in the context of pebble accretion, an important ingredient in recent theories of planet formation \citep[e.g.][for a review]{joha17}.
The \ce{H2O}-R$_{\rm{dust}}$ correlation suggests that retaining pebbles in the outer disk decreases the water content in the inner disk. Pebble drift may therefore be a major transport mechanism for water through the disk and inside the snow line \citep[as proposed for other molecules, e.g. CO;][]{krijt20}. If this is true, measurements of the infrared \ce{H2O} luminosity and/or of R$_{\rm{dust}}$ could be used to determine which disks are forming (or have formed) small rocky planets rather than super-Earths in the inner disk, which are proposed to depend on the mass-flux of migrating pebbles \citep{lamb19}.

Small disks with large water luminosity should have had a strong flux of pebbles delivering solid mass to form rocky planets within the snow line. These inner planets should be relatively dry, if pebbles release most of their ice content by sublimation. These disks also probably have not yet formed giant planets outside the snow line, or these planets would have prevented water from being delivered into the warm inner region \citep[following the interpretation by][]{naj13}. However, Super-Earths could still be easily formed inside the snow line in these disks, due to the large flux of migrating pebbles \citep{lamb19}.

Large disks with substructures and a low water luminosity, but without an inner cavity, are the best case supporting an inefficient pebble drift into the snow line. In other words, these disks have not delivered the same mass of ice to the region inside the snow line, compared to the small disks. This would imply that a large mass of icy pebbles are still retained in the disk at large radii and are available to form planets. Retaining pebbles in rings can make planet formation more difficult at larger disk radii, but still allow for fast planet formation around 5~au \citep{morbi20}. In these disks, formation of super-Earths inside the snow line can be hindered by the lower drift efficiency, leading to systems of small rocky planets \citep{lamb19}.

All these scenarios depend on the properties of the migrating icy pebbles (including size and composition) and how much oxygen mass they release per unit of pebble mass. It would be interesting in future work to study how detailed models of ice transport and sublimation might be able to match the observed trend between infrared water line luminosity and disk radius, and link them to the pebble mass that is delivered to the rocky planet-forming zone.

\subsection{Predictions for future work} \label{sec: future}
As discussed above, the current data and results find a natural interpretation in the context of a physical link between inner disk molecular abundances and the evolution of dust in disks, in terms of inward drift of icy pebbles and formation of inner cavities. It is clear that multiple and possibly interconnected processes affect the observed molecular luminosities, and this is at least in part the origin for the large scatters measured in the correlations analyzed in this work. In this section we discuss how this analysis can be improved in terms of data and samples, and we propose some fundamental predictions to test in future work.

In terms of infrared molecular spectra, the analysis can be improved in two main ways. The large wavelength coverage needed to characterize the emission, especially in the case of water, will be provided by the \textit{James Webb Space Telescope}-MIRI. JWST-MIRI spectra will help solve some extant problems of \textit{Spitzer} spectra (Section \ref{sec: data}). 
The factor $ \sim4$ higher spectral resolution will allow a better characterization of the local continuum and the de-blending of several (though not all) emission lines from different molecules and transitions \citep[e.g. Figure 5 in][]{pont10a}, allowing to isolate at least some optically thin lines that are important to measure column densities \citep[e.g. Chapter 4 in][]{banz13,notsu17}. The factor $\sim 10-100$ better sensitivity will allow to measure line fluxes down to weaker emission by a similar factor, and to better characterize all correlations especially where currently dominated by upper limits (in particular those with the infrared index $n_{13-30}$, see Section \ref{sec: n1330_corr}). The key products of these higher-resolution and higher-sensitivity observations, in the context of this analysis, will be to i) confirm whether the \ce{H2O}-R$_{\rm{dust}}$ relation corresponds to a similar relation between water column density or abundance and the disk radius (Appendix \ref{app: water_coldens}), and ii) determine any differences in water abundance between large disks with and without an inner cavity (Section \ref{sec: disc_cav} and Figure \ref{fig: cartoon}). 
Future mid-infrared spectral samples can also be designed to explore any dependence of the observed trends on other factors like environment and age, and their role in the evolution of inner disk molecular abundances. 

The second aspect of improvement will be in the collection of high-dispersion (R~$>30,000$) mid-infrared spectra of water and carbon-bearing molecules that allow the characterization of the spatial distribution of the emission from fits to the line profiles \citep[e.g.][]{salyk19}. Obtaining spatial information on the gas-emitting regions will allow to follow the depletion of gas as a function of disk radius, as done for CO, \ce{H2O}, and OH from high-dispersion (R\,$\sim100,000$) near-infrared spectra \citep[e.g.][]{bp15,banz17,bosm19}. This will be key also to study any difference in water vapor depletion between large disks with and without an inner cavity, which could happen in an homologous rather than radial fashion. Current estimates for near-infrared CO emission suggest that disks with inner cavities not only have a lower column density of gas as compared to ``full" disks but also a smaller emitting area \citep{salyk11b} from a narrower ring of emission at larger radii (Appendix \ref{app: water_coldens} and Figure \ref{fig: cartoon}).
While current ground-based facilities limit progress to small samples of bright disks \citep{salyk19}, the best solution to both spectral resolution and sensitivity requirements would be provided by a future space telescope like the \textit{Origins Space Telescope} \citep{pont18}.
In the meantime, infrared spectrographs on sub-orbital platforms could also be used to spectrally resolve (and therefore image with line tomography) water lines that trace the disk region near the water snow line \citep[e.g.][]{rich18}.

In terms of high-resolution imaging, there are at least a few ways to test and improve the current analysis.
While ALMA has directly imaged large-scale substructures in the outer region of large disks, the analysis of visibilities to study sub-resolution dust structures has hinted at smaller scale substructures existing in disks with R$_{\rm{dust}}$ as small as $\sim20$~au \citep{dsharp2,long20,kurt20}. Current data show that there is no preferred location for substructures and they might well be common in small disks too \citep{andr20}. Just as large-scale substructures in large disks may explain why pebbles are still present and detected at disk radii beyond 60~au, smaller scale substructures might be the reason why even smaller disks survive long enough to be observed at 1--3~Myr. If future work confirms substructures to be common in small disks, the fundamental prediction to test in the context of this analysis is that they have allowed for a larger mass-flux of icy pebbles to drift inside the water snow line, as compared to the mass-flux of drifting pebbles in large disks with large-scale substructures observed today. 

Moreover, as discussed above R$_{\rm{dust}}$ is currently the best proxy for dust drift efficiency, as supported by several observations and models (Section \ref{sec: disc_drift}). As of today, small disks have supported evidence for efficient radial drift of solid pebbles \citep[][]{trap19,trap20,facch19,kurt20}, confirming the interpretation we adopt in this work. However, a more direct proxy for dust drift efficiency would be the ratio of measured disk radii in gas and dust \citep[e.g.][]{trap20}, but gas disk radii from millimeter observations are still only sparsely available \citep{andr20}. When high S/N images of millimeter gas emission are obtained for larger samples of disks, they will provide a way to further test whether pebble drift is more efficient in small disks with large \ce{H2O} luminosity, by e.g. finding a larger R$_{\rm{gas}}$/R$_{\rm{dust}}$ ratio than in large disks.

Moving closer to the star, the inner 2--3\,au have yet to be resolved in most disks, to spatially-resolve the smallest inner dust structures. It is interesting to note that of the five disks with $n_{13-30} > 0$ and yet molecular emission detected, DoAr\,44 has been recently imaged with VLTI/GRAVITY spatially-resolving a residual inner dust belt at $\sim0.14$\,au \citep{bouv20}, which might explain why molecules have survived within the large inner dust cavity \citep{salyk15}. The other four disks with $n_{13-30} > 0$ and yet molecular emission detected (DH\,Tau, DoAr\,25, Haro\,6-13, IRAS 04385+2550) still lack high angular resolution observations to i) confirm the presence of a (small, possibly $\sim$~1--2-au-wide) inner dust cavity, and ii) investigate the presence of an inner dust-belt structure and confirm whether that is what is needed for molecules to survive in inner disk dust cavities.

\section{Summary and Conclusions} \label{sec: summary}
In this work, we have analyzed a sample of 63 T~Tauri disks where two types of data are available: \textit{i)} spatially-resolved disk images from millimeter interferometry with ALMA or the SMA, and \textit{ii)} molecular emission spectra as observed at mid-infrared wavelengths with \textit{Spitzer}. 
High-resolution millimeter imaging probes dust substructures and the radial distribution of disk pebbles (mm-cm dust grains), providing measurements of an effective outer disk radius R$_{\rm{dust}}$ \citep[e.g.][for a review]{andr20}. Mid-infrared spectra trace molecular gas in inner disks and the mass budgets of the most abundant elements \citep[e.g.][for a review]{pont14}. Building on a decade of analyses and on the current understanding of the relation between infrared spectra and stellar, accretion, and disk properties, we performed a systematic study of correlations between molecular luminosities for \ce{H2O}, HCN, \ce{C2H2}, and \ce{CO2}, R$_{\rm{dust}}$, and an infrared index that is sensitive to the presence an size of an inner disk cavity, n$_{13-30}$.

This analysis detects correlations between the flux ratio of water to the other molecules and R$_{\rm{dust}}$, expanding upon previous findings of a correlation between \ce{H2O}/HCN and dust disk mass \citep{naj13}. Normalization to a common dependence with the accretion luminosity further suggests that the strongest underlying relation is between \ce{H2O} and R$_{\rm{dust}}$, although the large measured scatters still allow shallower relations between the carbon-bearing molecules and R$_{\rm{dust}}$, which should be investigated in future work. If R$_{\rm{dust}}$ is mainly set by pebble drift rather than by different initial conditions, and if the molecular luminosities trace elemental mass budgets in inner disks, the results of this analysis find a natural explanation in a scenario where the inner disk molecular chemistry is fed by sublimation of water-rich icy pebbles that migrate to the inner disk, a fundamental prediction from decades ago that is now attracting increasing attention \citep[e.g.][]{cyr98,cc06,bosm18,booth19,krijt20}. After crossing the water snow line, the icy pebbles sublimate and enrich the inner disk with oxygen, thus lowering the C/O ratio and driving the efficient formation of water vapor (Section \ref{sec: disc_drift}).

While highly suggestive of a physical link between inner disk chemistry and outer disk evolution, the interpretation of these results still rely on key aspects that should be confirmed or corrected in future work. In particular, we highlight the following fundamental tests:
     
\textit{i)}  Future sensitive mid-infrared spectra (especially from \textit{JWST}-MIRI) will allow to improve current molecular flux upper limits by factors of $\sim$~10--100; analysis of these spectra should confirm whether the larger/lower \ce{H2O} luminosity in smaller/larger disks corresponds to an increase/decrease in inner disk \ce{H2O} abundance.
     
\textit{ii)}  Future sensitive surveys of disk radii in gas emission should confirm whether the measured R$_{\rm{dust}}$ at 1--3~Myr is primarily set by pebble drift, by finding large R$_{\rm{gas}}$/R$_{\rm{dust}}$ ratios, rather than set by different initial conditions when disks are born.
     
\textit{iii)}  If substructures are found to be common in small disks too, future work should confirm that these are not as efficient as the large-scale substructures observed today in preventing icy pebble drift from crossing the snow line and feeding oxygen to the inner disk chemistry.

A positive outcome from these tests will open a new exciting venue for synergic studies of the connections and causality between global disk evolution, the chemistry of planet-forming material, and the properties of exoplanet populations.

\acknowledgments
We thank the ALMA-Taurus team (2016.1.01164.S), Sean Andrews, Karin \"Oberg, and Michael Meyer for helpful discussions at the early stages of the development of this analysis, and the anonymous referee for helpful suggestions during revision. 
We also thank Nick Ballering for providing SED models used in the Appendix of this work. 
I.P. acknowledges support from a Collaborative NSF Astronomy \& Astrophysics Research grant (ID: 1715022). 
P.P. acknowledges support provided by the Alexander von Humboldt Foundation in the framework of the Sofja Kovalevskaja Award endowed by the Federal Ministry of Education and Research.
G.J.H. is supported by general grant 11773002 awarded by the National Science Foundation of China.

\facilities{ALMA, SMA, \textit{Spitzer}}

\software{{\tt linmix\_err} \citep{kelly07}}

\appendix

\section{Sample properties and measurements} \label{app: sample}
Tables \ref{tab: sample} and \ref{tab: fluxes} report the sample properties and line flux measurements, as described in Section \ref{sec: data}.

\begin{deluxetable*}{r l c c c c c c c}
\tabletypesize{\footnotesize}
\tablewidth{0pt}
\tablecaption{\label{tab: sample} Sample properties.}
\tablehead{\colhead{ID} & \colhead{Object name} & \colhead{Distance} & \colhead{M$_{\star}$} & \colhead{log L$_{\rm{acc}}$} & \colhead{log~R$_{\rm{dust}}$} & \colhead{n$_{13-30}$} & \colhead{R$_{\rm{cav}}$} & \colhead{References} 
\\ 
 &  & (pc) & (M$_{\odot}$) & (L$_{\odot}$) & (au) &  & (au) & }
\tablecolumns{9}
\startdata
       1 & 04385+2550$^a$ & 160.1 & 0.50 & -1.23 & 1.34 & 0.73 & -- & \textit{a2, r2} \\[\tblspc]
       2 & AA Tau      & 136.7 & 0.60 & -1.43 & 2.09 & -0.36 & 28. & \textit{a3, r1} \\[\tblspc]
       3 & AS 205 N     & 127.5 & 0.87 & -0.07 & 1.70 & -0.19 & -- & \textit{a1, r1} \\[\tblspc]
       4 & AS 209      & 120.6 & 0.96 & -1.12 & 2.14 & -0.28 & -- & \textit{a1, r1} \\[\tblspc]
       5 & BP Tau      & 128.6 & 0.54 & -1.17 & 1.62 & -0.36 & -- & \textit{a1, r1} \\[\tblspc]
       6 & CI Tau      & 158.0 & 0.71 & -0.87 & 2.28 & -0.17 & -- & \textit{a1, r1} \\[\tblspc]
       7 & CS Cha      & 175.4 & 0.74 & -1.31 & 1.74 & 2.96 & 37. & \textit{a4, r2} \\[\tblspc]
       8 & CV Cha      & 192.2 & 2.10 & 0.41 & 1.44 & -0.23 & -- & \textit{a3, r2} \\[\tblspc]
       9 & CX Tau      & 127.5 & 0.35 & -2.56 & 1.58 & -0.15 & -- & \textit{a1, r1} \\[\tblspc]
      10 & CY Tau      & 128.4 & 0.45 & -1.33 & 1.76 & -1.19 & -- & \textit{a2, r2} \\[\tblspc]
      11 & DH Tau      & 134.9 & 0.37 & -2.02 & 1.29 & 0.66 & -- & \textit{a1, r1} \\[\tblspc]
      12 & DK Tau      & 128.0 & 0.66 & -0.79 & 1.18 & -0.68 & -- & \textit{a1, r1} \\[\tblspc]
      13 & DL Tau      & 158.6 & 0.98 & -0.47 & 2.22 & -0.74 & -- & \textit{a3, r1} \\[\tblspc]
      14 & DM Tau      & 144.6 & 0.31 & -1.92 & 2.25 & 1.29 & 25. & \textit{a1, r2} \\[\tblspc]
      15 & DN Tau      & 127.8 & 0.55 & -1.93 & 1.78 & -0.13 & -- & \textit{a1, r1} \\[\tblspc]
      16 & DoAr 25     & 137.9 & 0.96 & -1.33 & 2.22 & 0.59 & -- & \textit{a2, r1} \\[\tblspc]
      17 & DoAr 44     & 145.3 & 1.22 & -0.73 & 1.89 & 0.80 & 34. & \textit{a1, r2} \\[\tblspc]
      18 & DO Tau      & 138.8 & 0.44 & -0.67 & 1.56 & -0.15 & -- & \textit{a2, r1} \\[\tblspc]
      19 & DQ Tau      & 196.4 & 1.61 & -- & 1.64 & -0.33 & -- & \textit{a3, r1} \\[\tblspc]
      20 & DR Tau      & 194.6 & 0.93 & -0.24 & 1.73 & -0.34 & -- & \textit{a3, r1} \\[\tblspc]
      21 & DS Tau      & 158.4 & 0.62 & -1.28 & 1.85 & -1.01 & -- & \textit{a1, r1} \\[\tblspc]
      22 & Elias 24    & 135.7 & 0.78 & 0.90 & 2.13 & -- & -- & \textit{a2, r1} \\[\tblspc]
      23 & FT Tau      & 127.3 & 0.34 & -1.11 & 1.66 & -0.34 & -- & \textit{a3, r1} \\[\tblspc]
      24 & GI Tau      & 130.0 & 0.53 & -0.69 & 1.39 & -0.79 & -- & \textit{a1, r1} \\[\tblspc]
      25 & GK Tau      & 128.8 & 0.67 & -1.38 & 1.11 & -0.37 & -- & \textit{a1, r1} \\[\tblspc]
      26 & GM Aur      & 159.0 & 1.32 & -1.15 & 2.27 & 1.75 & 34. & \textit{a2, r2} \\[\tblspc]
      27 & GQ Lup      & 151.2 & 0.78 & -0.36 & 1.34 & -0.18 & -- & \textit{a1, r2} \\[\tblspc]
      28 & GW Lup      & 155.2 & 0.37 & -1.87 & 2.02 & -0.22 & -- & \textit{a1, r1} \\[\tblspc]
      29 & Haro 6-13   & 130.0 & 0.55 & -0.40 & 1.54 & 0.67 & -- & \textit{a2, r1} \\[\tblspc]
      30 & HD 135344 B  & 135.3 & 1.43 & -1.11 & 1.98 & 1.85 & 62. & \textit{a3, r2} \\[\tblspc]
      31 & HD 143006   & 165.5 & 1.52 & -0.66 & 1.91 & 1.20 & 6. & \textit{a1, r1} \\[\tblspc]
      32 & HK Tau      & 132.9 & 0.44 & -- & 1.46 & 1.03 & -- & \textit{a3, r1} \\[\tblspc]
      33 & HN Tau      & 136.1 & 0.69 & -0.93 & 1.27 & -0.62 & -- & \textit{a1, r1} \\[\tblspc]
      34 & HQ Tau      & 158.2 & 1.78 & -1.60 & 1.39 & -0.50 & -- & \textit{a1, r1} \\[\tblspc]
      35 & HT Lup      & 153.5 & 1.27 & -1.18 & 1.40 & -0.40 & -- & \textit{a1, r1} \\[\tblspc]
      36 & IM Lup      & 157.7 & 0.67 & -1.75 & 2.42 & -0.30 & -- & \textit{a1, r1} \\[\tblspc]
      37 & IP Tau      & 130.1 & 0.59 & -2.29 & 1.56 & 0.14 & 35. & \textit{a1, r1} \\[\tblspc]
      38 & IQ Tau      & 130.8 & 0.50 & -- & 2.04 & -0.37 & -- & \textit{a2, r1} \\[\tblspc]
      39 & LkCa 15     & 158.2 & 0.76 & -1.70 & 2.20 & 0.53 & 48. & \textit{a1, r2} \\[\tblspc]
      40 & LkHa 330    & 308.4 & 2.95 & -0.46 & 2.26 & 1.88 & 68. & \textit{a2, r2} \\[\tblspc]
      41 & MY Lup      & 155.9 & 1.23 & -0.70 & 1.94 & 0.19 & -- & \textit{a5, r1} \\[\tblspc]
      42 & RU Lup      & 158.9 & 0.55 & -0.01 & 1.80 & -0.53 & -- & \textit{a1, r1} \\[\tblspc]
      43 & RW Aur      & 163.0 & 1.20 & -0.20 & 1.33 & -0.52 & -- & \textit{a5, r1} \\[\tblspc]
      44 & RY Lup      & 158.4 & 1.27 & -1.40 & 2.09 & 0.87 & 68. & \textit{a1, r2} \\[\tblspc]
      45 & RY Tau      & 128.0 & 2.04 & 0.07 & 1.81 & -0.19 & 21. & \textit{a3, r1} \\[\tblspc]
      46 & SR 4        & 134.1 & 0.68 & 0.06 & 1.49 & 0.48 & -- & \textit{a2, r1} \\[\tblspc]
      47 & SR 21       & 137.9 & 1.79 & -0.70 & 1.88 & 2.01 & 51. & \textit{a4, r2} \\[\tblspc]
      48 & SX Cha      & 184.0 & 0.77 & -1.25 & 1.40 & -0.48 & -- & \textit{a4, r2} \\[\tblspc]
      49 & SY Cha      & 182.1 & 0.78 & -2.24 & 2.27 & -0.17 & 45. & \textit{a4, r2} \\[\tblspc]
      50 & Sz 73       & 156.1 & 0.75 & -1.22 & 1.54 & -0.06 & -- & \textit{a1, r2} \\[\tblspc]
      51 & TW Cha      & 184.2 & 1.00 & -1.66 & 1.72 & -0.16 & -- & \textit{a4, r2} \\[\tblspc]
      52 & TW Hya      & 60.0 & 0.61 & -1.53 & 1.76 & 0.96 & 2. & \textit{a1, r2} \\[\tblspc]
      53 & UY Aur      & 155.0 & 0.65 & -1.13 & 0.83 & -0.03 & -- & \textit{a3, r1} \\[\tblspc]
      54 & UX Tau A     & 139.4 & 1.40 & -1.51 & 1.69 & 1.82 & 38. & \textit{a1, r2} \\[\tblspc]
      55 & V710 Tau    & 142.0 & 0.42 & -- & 1.65 & -0.32 & -- & \textit{a3, r1} \\[\tblspc]
      56 & V836 Tau    & 168.8 & 0.72 & -2.51 & 1.50 & -0.07 & -- & \textit{a1, r1} \\[\tblspc]
      57 & V853 Oph    & 137.0 & 0.32 & -1.46 & 1.75 & -0.17 & -- & \textit{a1, r2} \\[\tblspc]
      58 & VSSG 1      & 138.1 & 0.48 & -0.87 & 1.81 & -0.36 & -- & \textit{a2, r1} \\[\tblspc]
      59 & VW Cha      & 190.0 & 0.60 & -0.77 & 1.32 & -0.13 & -- & \textit{a3, r2} \\[\tblspc]
      60 & VZ Cha      & 191.2 & 0.80 & -0.73 & 1.60 & -1.08 & -- & \textit{a3, r2} \\[\tblspc]
      61 & Wa Oph 6     & 123.4 & 0.63 & -0.66 & 2.01 & -0.40 & -- & \textit{a1, r1} \\[\tblspc]
      62 & WX Cha      & 190.0 & 0.50 & -0.83 & 1.25 & -1.03 & -- & \textit{a5, r2} \\[\tblspc]
      63 & XX Cha      & 189.5 & 0.25 & -0.81 & 1.31 & -0.26 & -- & \textit{a4, r2} \\
\enddata

\tablecomments{
Distances are taken from \citet{bj18}. M$_{\star}$ and L$_{\rm{acc}}$ references: \textit{a1} \citet{herc14,fang18,sim16}; \textit{a2} \citet{andr18a,dsharp1}; \textit{a3} \citet{long19,salyk13}; \textit{a4} \citet{man14,man16}; \textit{a5} \citet{alc19,cos12,whi01}. R$_{\rm{dust}}$ references (see Section \ref{sec: radii} for details): \textit{r1} \citet{dsharp2}, \citet{kurt18}, \citet{long18,long19}, \citet{facch19}, \citet{pin20}; \textit{r2} \citet{andr18a} and \citet{hend20}. Infrared index n$_{13-30}$ values are measured from \textit{Spitzer} spectra (Section \ref{sec: data}). Millimeter cavity radii R$_{\rm{cav}}$ are adopted from \citet{andr16,pasc16,loom17,mac18,dsharp2,long18,pin18,fran20}. $^a$IRAS 04385+2550.
}
\end{deluxetable*}

\begin{deluxetable*}{r l c c c c}
\tabletypesize{\footnotesize}
\tablewidth{0pt}
\tablecaption{\label{tab: fluxes} Molecular line flux measurements.}
\tablehead{\colhead{ID} & \colhead{Object name} & \colhead{\ce{H2O} flux} & \colhead{HCN flux} & \colhead{\ce{C2H2} flux} & \colhead{\ce{CO2} flux}
\\ 
 &  & \multicolumn{4}{|c|}{(10$^{-14}$ erg s$^{-1}$ cm$^{-2}$)}  }
\tablecolumns{6}
\startdata
       1 & 04385+2550$^a$ & 3.33 $\pm$ 0.17 & 3.10 $\pm$ 0.11 & 1.32 $\pm$ 0.08 & 1.81 $\pm$ 0.08 \\[\tblspc]
       2 & AA Tau      & 4.68 $\pm$ 0.15 & 7.00 $\pm$ 0.07 & 2.23 $\pm$ 0.05 & 1.04 $\pm$ 0.07 \\[\tblspc]
       3 & AS 205 N     & 53.77 $\pm$ 2.23 & 39.04 $\pm$ 1.12 & 18.99 $\pm$ 0.87 & 20.06 $\pm$ 0.72 \\[\tblspc]
       4 & AS 209      & (5.03 $\pm$ 3.39) & (0.16 $\pm$ 3.00) & (2.13 $\pm$ 2.65) & 4.94 $\pm$ 0.40 \\[\tblspc]
       5 & BP Tau      & 6.55 $\pm$ 0.13 & 6.12 $\pm$ 0.10 & 0.90 $\pm$ 0.07 & 0.78 $\pm$ 0.04 \\[\tblspc]
       6 & CI Tau      & 5.64 $\pm$ 0.12 & 5.79 $\pm$ 0.10 & 2.75 $\pm$ 0.07 & 1.51 $\pm$ 0.06 \\[\tblspc]
       7 & CS Cha      & (-0.19 $\pm$ 0.62) & (-0.37 $\pm$ 0.68) & (-0.12 $\pm$ 0.86) & (0.00 $\pm$ 0.16) \\[\tblspc]
       8 & CV Cha      & 16.51 $\pm$ 2.21 & (3.59 $\pm$ 5.51) & (0.50 $\pm$ 2.25) & (0.75 $\pm$ 1.32) \\[\tblspc]
       9 & CX Tau      & (0.98 $\pm$ 1.50) & 1.41 $\pm$ 0.17 & (0.25 $\pm$ 0.33) & (1.18 $\pm$ 1.53) \\[\tblspc]
      10 & CY Tau      & 0.99 $\pm$ 0.07 & 0.87 $\pm$ 0.04 & 1.54 $\pm$ 0.03 & 0.38 $\pm$ 0.03 \\[\tblspc]
      11 & DH Tau      & 2.36 $\pm$ 0.18 & (0.39 $\pm$ 0.44) & (0.12 $\pm$ 0.16) & (0.15 $\pm$ 0.32) \\[\tblspc]
      12 & DK Tau      & 15.28 $\pm$ 0.21 & 7.00 $\pm$ 0.18 & 1.28 $\pm$ 0.10 & 1.56 $\pm$ 0.10 \\[\tblspc]
      13 & DL Tau      & 2.29 $\pm$ 0.16 & 4.55 $\pm$ 0.10 & 6.02 $\pm$ 0.08 & 0.52 $\pm$ 0.06 \\[\tblspc]
      14 & DM Tau      & (0.37 $\pm$ 0.26) & (0.23 $\pm$ 0.27) & (0.12 $\pm$ 0.15) & (0.12 $\pm$ 0.13) \\[\tblspc]
      15 & DN Tau      & (0.96 $\pm$ 0.95) & 2.21 $\pm$ 0.24 & 0.89 $\pm$ 0.18 & 1.19 $\pm$ 0.16 \\[\tblspc]
      16 & DoAr 25     & (1.49 $\pm$ 0.68) & 4.71 $\pm$ 0.22 & 0.88 $\pm$ 0.16 & 0.59 $\pm$ 0.15 \\[\tblspc]
      17 & DoAr 44     & 4.82 $\pm$ 0.16 & 3.26 $\pm$ 0.08 & (0.59 $\pm$ 0.65) & (0.30 $\pm$ 0.38) \\[\tblspc]
      18 & DO Tau      & 4.74 $\pm$ 0.25 & 5.30 $\pm$ 0.18 & 2.97 $\pm$ 0.12 & 1.36 $\pm$ 0.12 \\[\tblspc]
      19 & DQ Tau      & 4.71 $\pm$ 0.30 & 3.76 $\pm$ 0.18 & 0.93 $\pm$ 0.15 & 2.65 $\pm$ 0.26 \\[\tblspc]
      20 & DR Tau      & 23.15 $\pm$ 0.48 & 17.84 $\pm$ 0.35 & 4.08 $\pm$ 0.25 & 4.22 $\pm$ 0.24 \\[\tblspc]
      21 & DS Tau      & 4.03 $\pm$ 0.25 & 7.48 $\pm$ 0.19 & 1.50 $\pm$ 0.13 & 0.88 $\pm$ 0.08 \\[\tblspc]
      22 & Elias 24    & 25.78 $\pm$ 1.71 & 21.30 $\pm$ 1.95 & (6.24 $\pm$ 20.52) & 4.51 $\pm$ 0.60 \\[\tblspc]
      23 & FT Tau      & 1.96 $\pm$ 0.09 & 2.51 $\pm$ 0.08 & 0.94 $\pm$ 0.05 & 0.64 $\pm$ 0.03 \\[\tblspc]
      24 & GI Tau      & 6.31 $\pm$ 0.19 & 6.74 $\pm$ 0.17 & 2.66 $\pm$ 0.11 & 0.78 $\pm$ 0.10 \\[\tblspc]
      25 & GK Tau      & 4.08 $\pm$ 0.22 & 3.30 $\pm$ 0.09 & 1.47 $\pm$ 0.07 & 0.54 $\pm$ 0.10 \\[\tblspc]
      26 & GM Aur      & (0.35 $\pm$ 0.53) & (0.74 $\pm$ 0.79) & (0.10 $\pm$ 0.35) & (0.40 $\pm$ 0.53) \\[\tblspc]
      27 & GQ Lup      & 4.76 $\pm$ 0.14 & 3.70 $\pm$ 0.06 & 0.97 $\pm$ 0.05 & 0.75 $\pm$ 0.05 \\[\tblspc]
      28 & GW Lup      & (0.64 $\pm$ 0.86) & (1.04 $\pm$ 1.06) & 0.69 $\pm$ 0.16 & 0.89 $\pm$ 0.16 \\[\tblspc]
      29 & Haro 6-13   & 4.74 $\pm$ 0.27 & 4.64 $\pm$ 0.16 & 1.35 $\pm$ 0.13 & 3.20 $\pm$ 0.13 \\[\tblspc]
      30 & HD 135344 B  & (0.33 $\pm$ 1.53) & (-3.01 $\pm$ 1.25) & (-1.91 $\pm$ 0.70) & (-0.37 $\pm$ 0.61) \\[\tblspc]
      31 & HD 143006   & (0.23 $\pm$ 1.46) & (-2.10 $\pm$ 1.03) & (-0.32 $\pm$ 0.57) & (-1.15 $\pm$ 0.52) \\[\tblspc]
      32 & HK Tau      & (0.74 $\pm$ 0.85) & (0.08 $\pm$ 0.20) & 0.22 $\pm$ 0.03 & 0.61 $\pm$ 0.03 \\[\tblspc]
      33 & HN Tau      & 3.65 $\pm$ 0.20 & 2.58 $\pm$ 0.18 & 1.22 $\pm$ 0.10 & 0.60 $\pm$ 0.13 \\[\tblspc]
      34 & HQ Tau      & (0.24 $\pm$ 0.58) & 1.67 $\pm$ 0.17 & 0.77 $\pm$ 0.11 & (0.34 $\pm$ 0.43) \\[\tblspc]
      35 & HT Lup      & 4.84 $\pm$ 1.06 & 10.34 $\pm$ 0.50 & (0.88 $\pm$ 1.34) & 4.21 $\pm$ 0.38 \\[\tblspc]
      36 & IM Lup      & (0.65 $\pm$ 0.89) & 1.27 $\pm$ 0.06 & 1.39 $\pm$ 0.05 & 1.15 $\pm$ 0.05 \\[\tblspc]
      37 & IP Tau      & (0.74 $\pm$ 0.83) & (0.56 $\pm$ 1.02) & (-0.06 $\pm$ 0.57) & (-0.07 $\pm$ 0.51) \\[\tblspc]
      38 & IQ Tau      & 3.26 $\pm$ 0.34 & 7.12 $\pm$ 0.27 & 2.77 $\pm$ 0.23 & (0.63 $\pm$ 0.70) \\[\tblspc]
      39 & LkCa 15     & (0.92 $\pm$ 1.35) & (-1.14 $\pm$ 0.94) & (-0.64 $\pm$ 1.14) & (-0.03 $\pm$ 1.08) \\[\tblspc]
      40 & LkHa 330    & (0.42 $\pm$ 0.51) & (0.08 $\pm$ 0.35) & (-0.14 $\pm$ 0.16) & (-0.29 $\pm$ 0.11) \\[\tblspc]
      41 & MY Lup      & (0.29 $\pm$ 1.16) & (0.65 $\pm$ 1.83) & (-0.35 $\pm$ 0.75) & 1.20 $\pm$ 0.15 \\[\tblspc]
      42 & RU Lup      & 12.15 $\pm$ 0.34 & 10.57 $\pm$ 0.31 & 3.84 $\pm$ 0.18 & 4.83 $\pm$ 0.17 \\[\tblspc]
      43 & RW Aur      & 22.62 $\pm$ 0.36 & 11.66 $\pm$ 0.36 & 5.88 $\pm$ 0.21 & 6.15 $\pm$ 0.22 \\[\tblspc]
      44 & RY Lup      & (-0.69 $\pm$ 1.97) & (-1.79 $\pm$ 2.01) & (-0.35 $\pm$ 1.13) & (-0.82 $\pm$ 0.98) \\[\tblspc]
      45 & RY Tau      & (7.88 $\pm$ 57.60) & (4.86 $\pm$ 19.97) & (5.87 $\pm$ 63.90) & (3.02 $\pm$ 68.08) \\[\tblspc]
      46 & SR 4        & (6.23 $\pm$ 9.12) & (4.09 $\pm$ 8.20) & (3.29 $\pm$ 7.05) & (0.74 $\pm$ 1.02) \\[\tblspc]
      47 & SR 21       & (-3.06 $\pm$ 3.59) & (-18.75 $\pm$ 2.89) & (-7.54 $\pm$ 1.63) & 2.12 $\pm$ 0.51 \\[\tblspc]
      48 & SX Cha      & 3.56 $\pm$ 0.14 & 0.60 $\pm$ 0.13 & (-0.05 $\pm$ 0.22) & (0.12 $\pm$ 0.21) \\[\tblspc]
      49 & SY Cha      & 0.61 $\pm$ 0.06 & 1.13 $\pm$ 0.03 & 0.59 $\pm$ 0.02 & 0.48 $\pm$ 0.02 \\[\tblspc]
      50 & Sz 73       & (2.07 $\pm$ 3.10) & (-2.40 $\pm$ 2.54) & (-0.18 $\pm$ 3.46) & 0.95 $\pm$ 0.31 \\[\tblspc]
      51 & TW Cha      & 3.63 $\pm$ 0.05 & 2.52 $\pm$ 0.03 & 0.85 $\pm$ 0.03 & (0.23 $\pm$ 0.26) \\[\tblspc]
      52 & TW Hya      & (0.39 $\pm$ 0.37) & (0.47 $\pm$ 0.52) & (0.70 $\pm$ 0.74) & 0.50 $\pm$ 0.05 \\[\tblspc]
      53 & UY Aur      & (6.40 $\pm$ 7.79) & (10.80 $\pm$ 14.89) & (3.41 $\pm$ 8.19) & 4.82 $\pm$ 0.99 \\[\tblspc]
      54 & UX Tau A     & (-0.63 $\pm$ 1.09) & (-0.49 $\pm$ 0.86) & (-0.50 $\pm$ 1.15) & (-0.09 $\pm$ 0.20) \\[\tblspc]
      55 & V710 Tau    & (0.93 $\pm$ 0.58) & 3.78 $\pm$ 0.22 & 1.72 $\pm$ 0.15 & (0.36 $\pm$ 0.53) \\[\tblspc]
      56 & V836 Tau    & 1.47 $\pm$ 0.15 & (0.66 $\pm$ 0.92) & (-0.05 $\pm$ 0.38) & 0.24 $\pm$ 0.07 \\[\tblspc]
      57 & V853 Oph    & (1.36 $\pm$ 0.72) & 1.99 $\pm$ 0.24 & 1.48 $\pm$ 0.18 & (0.18 $\pm$ 0.59) \\[\tblspc]
      58 & VSSG 1      & 8.97 $\pm$ 0.80 & 13.59 $\pm$ 0.40 & 10.48 $\pm$ 0.31 & 2.35 $\pm$ 0.33 \\[\tblspc]
      59 & VW Cha      & 11.48 $\pm$ 0.19 & 5.65 $\pm$ 0.08 & 1.27 $\pm$ 0.06 & 2.08 $\pm$ 0.04 \\[\tblspc]
      60 & VZ Cha      & 4.48 $\pm$ 0.12 & 5.22 $\pm$ 0.08 & 3.12 $\pm$ 0.06 & 0.63 $\pm$ 0.03 \\[\tblspc]
      61 & Wa Oph 6     & 5.38 $\pm$ 0.17 & 2.78 $\pm$ 0.11 & 1.11 $\pm$ 0.08 & 2.65 $\pm$ 0.10 \\[\tblspc]
      62 & WX Cha      & 5.28 $\pm$ 0.09 & 5.06 $\pm$ 0.10 & 1.61 $\pm$ 0.05 & 1.06 $\pm$ 0.05 \\[\tblspc]
      63 & XX Cha      & 1.66 $\pm$ 0.08 & 1.47 $\pm$ 0.05 & 1.26 $\pm$ 0.03 & 0.74 $\pm$ 0.03 \\
      \enddata

\tablecomments{
Molecular line fluxes are measured as explained in Section \ref{sec: data}. In this table, we report the carbon-bearing molecular line fluxes as corrected for water contamination. Fluxes not detected at $> 3 \sigma$ are reported in parentheses. $^a$IRAS 04385+2550.
}
\end{deluxetable*}

\section{Additional correlation tests} \label{app: more_correl}
Table \ref{tab: correlations_appendix} reports results from the Pearson and Spearman correlation tests for comparison to those using the method by \citet{kelly07} reported in Table \ref{tab: correlations}.

\begin{deluxetable*}{l c c c c c c}
\tabletypesize{\footnotesize}
\tablewidth{0pt}
\tablecaption{\label{tab: correlations_appendix} Additional correlation tests.}
\tablehead{\colhead{ } & \multicolumn{2}{c}{log L$_{\rm{acc}}$ (L$_\odot$)} & \multicolumn{2}{c}{log R$_{\rm{dust}}$ (au)} & \multicolumn{2}{c}{n$_{13-30}$}
\\ 
 & Pearson & Spearman  & Pearson & Spearman  & Pearson & Spearman }
\tablecolumns{7}
\startdata
\ce{H2O} & \textbf{0.72 ($<$0.1)} & \textbf{0.62 ($<$0.1)} & -0.19 (25) & -0.15 (39) & -0.12 (47) & -0.14 (41) \\
\ce{HCN} & \textbf{0.68 ($<$0.1)} & \textbf{0.65 ($<$0.1)} & -0.01 (99) & 0.01 (96) & -0.07 (66) & -0.22 (19) \\
\ce{C2H2} & \textbf{0.63 ($<$0.1)} & \textbf{0.57 ($<$0.1)} & 0.08 (64) & 0.04 (82) & \textbf{-0.39 (1.7)} & -0.29 (7.9) \\
\ce{CO2} & \textbf{0.57 ($<$0.1)} & \textbf{0.57 ($<$0.1)} & -0.07 (65) & -0.03 (87) & -0.07 (68) & 0.03 (84) \\
\ce{H2O}/HCN & 0.22 (19) & 0.28 (13) & \textbf{-0.50 (0.3)} & \textbf{-0.45 (0.7)} & -0.05 (80) & 0.05 (79) \\
\ce{H2O}/\ce{C2H2} & 0.22 (26) & 0.30 (12) & \textbf{-0.49 (0.6)} & \textbf{-0.46 (1.1)} & 0.15 (41) & 0.12 (51) \\
\ce{H2O}/\ce{CO2} & 0.11 (57) & 0.01 (96) & -0.26 (16) & -0.32 (8.3) & \textbf{-0.41 (2.6)} & -0.32 (8.7) \\
\ce{H2O}/L$^{0.6}_{\rm{acc}}$ & -- & -- & -0.27 (12) & -0.25 (15) & 0.02 (92) & 0.11 (54) \\
HCN/L$^{0.6}_{\rm{acc}}$ & -- & -- & 0.08 (63) & 0.15 (38) & -0.01 (95) & 0.14 (41) \\
\ce{C2H2}/L$^{0.6}_{\rm{acc}}$ & -- & -- & 0.26 (15) & 0.37 (8.6) & -0.15 (40) & 0.04 (82) \\
\ce{CO2}/L$^{0.6}_{\rm{acc}}$ & -- & -- & 0.01 (99) & 0.09 (60) & 0.06 (73) & 0.24 (14) \\
\enddata

\tablecomments{
Correlation coefficients and percent p-values (in brackets) for the Pearson and Spearman correlation tests for comparison to Table \ref{tab: correlations} (see Section \ref{sec: results}); correlations that are considered detected and significant are reported in boldface. 
}
\end{deluxetable*}

\section{The impact of wide binaries on the \ce{H2O}-R$_{\rm{dust}}$ relation} \label{app: binaries}

\begin{figure}
\centering
\includegraphics[width=0.5\textwidth]{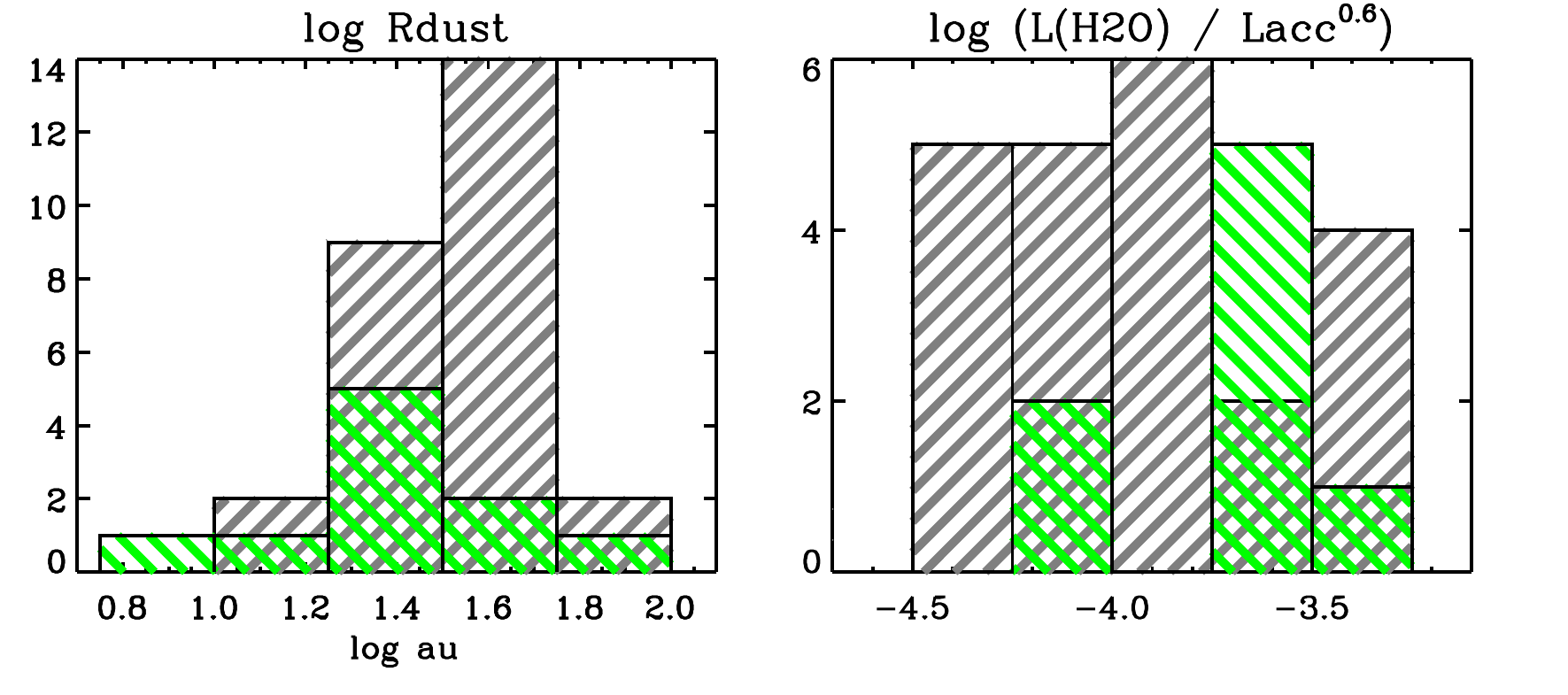} 
\caption{Distribution of binary systems (in green) as compared to single stars (grey) in the same range of disk radii (R$_{dust} < 60$~au).}
\label{fig: binary}
\end{figure}

The sample includes 10 known binary systems with separations $\sim$~0.1"--3.5" \citep{dsharp2,long19,schae18}: AS~205, DH~Tau, DK~Tau, HK~Tau, HN~Tau, HT~Lup, RW~Aur, UY~Aur, V710~Tau, V853~Oph. In all these cases, the \textit{Spitzer}-IRS spectra include both A and B components on the slit (the IRS SH slit width is 4.7"). The primary (A) components dominate the observed flux at millimeter as well as infrared wavelengths \citep{dsharp2,long19,naj13}, so that the molecular spectra used in this work are associated to the dust disk radii of the primaries (Table \ref{tab: sample}). 
Figure \ref{fig: binary} shows how the binary systems in this sample locate in terms of disk radii and \ce{H2O} luminosity as compared to disks of similar sizes (R$_{dust} < 60$~au) around single stars. Even if disks in binaries on average show smaller dust radii than disks around single stars \citep{long19,man19}, the two distributions overlap and in this sample single stars outnumber the binaries. It is interesting to note that binaries mix very well with single stars in terms of the \ce{H2O}-R$_{\rm{dust}}$ relation, suggesting high columns of inner disk water vapor as a result of an efficient pebble drift (Section \ref{sec: disc_drift}).
In fact, we note that the interpretation of small disks in the context of pebble drift seems to be true also in wide binary systems, because the observed dust radii are too small to be explained by truncation by a companion and pebble drift could be even more efficient than in disks around single stars \citep[][]{man19}.

\begin{figure*}
\centering
\includegraphics[width=1\textwidth]{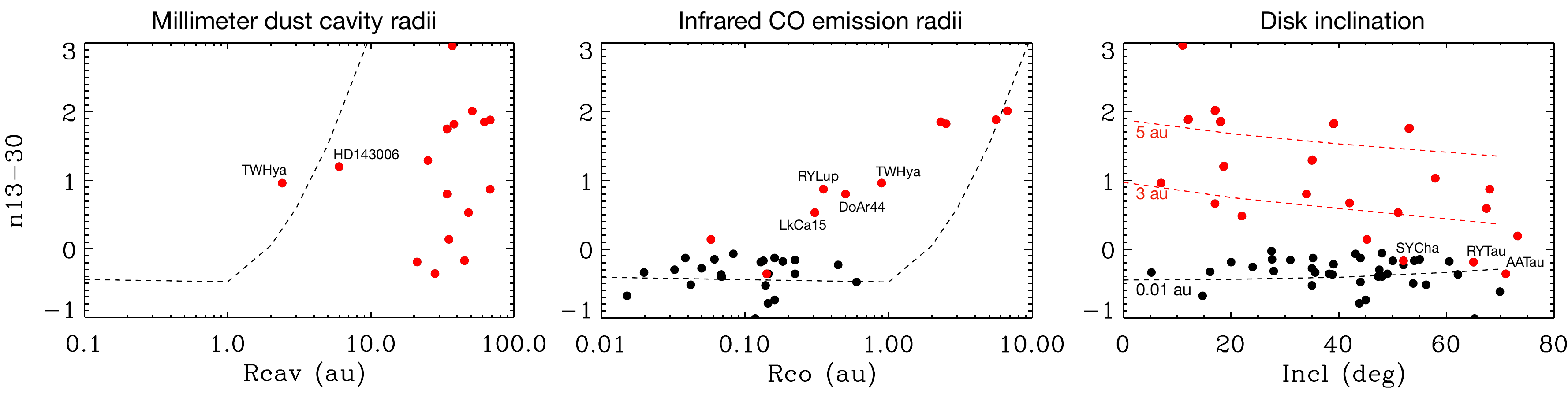} 
\caption{Infrared index n$_{13-30}$ as a function of inner disk cavity size and disk inclination. The dashed lines show models by \citet{ball19}. Left: using the dust cavity radius from millimeter observations (Table \ref{tab: sample}). Middle: using the radius of infrared CO emission R$_{\rm{co}}$ from \citet{banz17} as a proxy for an inner cavity size. Right: using disk inclinations; the inner disk cavity size for each model is shown next to each line. Red datapoints identify disks with an inner dust cavity.}
\label{fig: n1330_Rcav}
\end{figure*}

\section{n$_{13-30}$ as a probe of inner disk cavity size} \label{app: IRindex}
The infrared index n$_{13-30}$ was originally studied by \citet{furl06,furl09} in the context of dust settling in disk surfaces, based on models by \citet{dal06}. These models proposed that $0<$~n$_{13-30} < 1$ should be found in disks where small and large dust grains are well mixed in the disk upper layers, while settling of large grains toward disk midplanes will produce n$_{13-30} < 0$ (for typical accretion rates of $10^{-8}$~M$_{\odot}$/yr). The prevalence of disks with n$_{13-30} < 0$ led the authors to conclude for widespread evidence for dust settling in disk atmospheres. Later, \citet{brow07} introduced n$_{13-30}$ as a diagnostic of ``cold disks", i.e. disks with deficit of emission from warm dust due to large inner gaps in their radial distribution of dust grains. From the distribution of n$_{13-30}$ values in their sample, the authors identified ``transitional" disks to be in the range of $0.9<$~n$_{13-30} < 2.2$ \citep[or a flux ratio of 5--15 in Figure 1 of][]{brow07}, a range later adopted also by \citet{furl09}. \citet{salyk09} adopted the same diagnostic and lowered the value to identify ``transitional" disks to n$_{13-30} > 0.3$, based on their sample. In fact, going back to the works by \citet{furl06,furl09} shows that although models formally allowed n$_{13-30}$ as high as 1 in disks without inner cavities, the data actually showed that n$_{13-30} > 0.3$ was found only in outliers and those disks that at the time had been identified as ``transitional". While there is to date no clear cut in n$_{13-30}$ to identify disks with an inner cavity, we offer in the following a comparison between the data collected for this sample and recent models available in the literature.

Figure \ref{fig: n1330_Rcav} shows the data used in this work in comparison to models of a disk around a T~Tauri star. We adopt the fiducial model from \citet{ball19}, here explored over a grid of inner disk cavity radii up to 10\,au and over a range of disk inclinations. 
The models show a strong increase in n$_{13-30}$ for dust cavity radii larger than 1\,au. These models are clearly just an approximation, as they assume a fully-devoid inner cavity without a sophisticated treatment of the region around the inner cavity wall, so we use them here only as a first step to investigate relations between n$_{13-30}$ and the size of an inner disk cavity.
In terms of data, it is not trivial to identify a tracer of the size of inner disk cavities, as these are known to depend on the wavelength of observations and their dependence of dust grain sizes \citep[e.g.][]{gar13}. Dust cavity radii from spatially-resolved millimeter observations (Figure \ref{fig: n1330_Rcav}, left) generally do not probe the inner location of hot dust that dominates the observed mid-infrared flux and the n$_{13-30}$ index. The only two millimeter cavities that are consistent with the model are those detected in TW~Hya and HD~143006, which are among the few observed at the highest spatial resolution achievable with ALMA.
The radius of infrared CO emission R$_{\rm{co}}$ (Figure \ref{fig: n1330_Rcav}, middle) generally seems to probe closely the size of an inner region depleted in dust, as proposed in previous work \citep{bp15,banz17,banz18,bosm19,anton20}. The trend provided by the model is well matched by most datapoints, but some disks with inner cavities clearly deviate from the model (these are marked in the figure). In these cases, the infrared molecular emission most likely probes a residual dense inner dust ring inside a larger dust cavity \citep{salyk15}, which has been detected or suggested in at least three of these disks \citep[LkCa~15, DoAr~44 and RY~Lup, see][]{mans09,talm16,aru18,bouv20}.

\begin{figure*}
\centering
\includegraphics[width=1\textwidth]{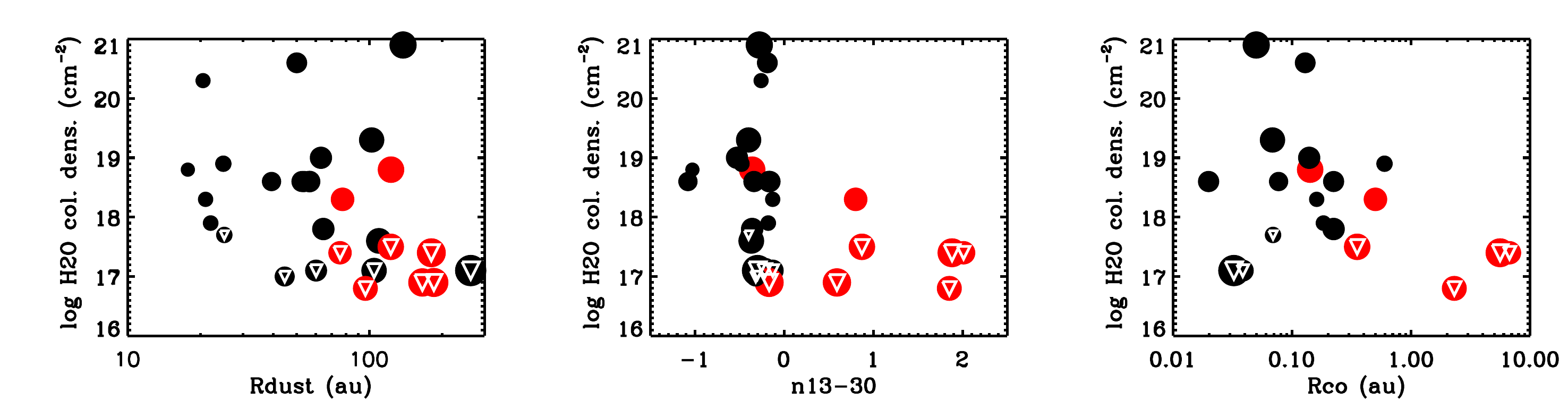} 
\caption{Current estimates of inner disk warm water column densities from \citet{salyk11a}. Colors follow the same code as in the rest of the paper. To help read the multi-dimensional dependence of \ce{H2O} column densities, the symbol size is proportional to R$_{\rm{dust}}$. Upper limits are marked with white triangles.}
\label{fig: S11_fits}
\end{figure*}

It is also interesting to note the dependence of n$_{13-30}$ on viewing geometry (Figure \ref{fig: n1330_Rcav}, right). Even here, simple inner disk models overall match pretty well the difference between disks that do not have an inner dust cavity versus those that have one. In disks with inner cavities, the data show a clear overall trend with n$_{13-30}$ decreasing in more inclined disks. Models reproduce this trend, although possibly with a somewhat shallower slope \citep[again, bear in mind that no attempt to match the data was done, here we are taking the models as published in ][]{ball19}. The interesting implication of this analysis is that inner dust cavities can be ``hidden" in highly-inclined disks. The only three disks with n$_{13-30} < 0$ and yet a spatially-resolved millimeter dust cavity all have high disk inclinations (where available, we take the inner disk inclination rather than the outer disk inclination): AA~Tau \citep[incl = 71 deg,][]{cox13}, RY~Tau \citep[incl = 65 deg,][]{long18}, and SY~Cha \citep[incl = 52 deg,][]{hend20}.
We conclude therefore that n$_{13-30}$ should overall be a good tracer of the presence and size of an inner disk dust cavity (with the caveat of highly-inclined disks), and that this index can be especially useful to reveal small inner cavities and/or cavities in disks that are too distant to be spatially resolved with ALMA. However, this part of the SED is clearly affected to a lower level by a number of other disk properties \citep[e.g.][]{furl09,woitke16,ball19}, so that interpreting n$_{13-30}$ for individual disks surely needs detailed modeling. But for the analyses of large disk samples, it would be interesting in future work to study the relation of n$_{13-30}$ to different inner disk structures, beyond the early studies focused on global dust settling without inner cavities or gaps \citep[e.g.][]{furl09}, to clarify the interpretation of correlations that are being found with inner disk gas tracers \citep[][and this work]{salyk11a,banz19,pasc20}.

\section{Current estimates of warm water columns in inner disks} \label{app: water_coldens}

Figure \ref{fig: S11_fits} shows warm water column densities estimated in \citet{salyk11a} from \textit{Spitzer} spectra, together with the data used in this work. Although these estimates are known to only approximately capture the complexity of inner disks, by assuming a single slab of gas with one set of area, temperature, and column density, they have been found to generally reproduce water emission in \textit{Spitzer} spectra \citep{salyk11a,cn11}. These mid-infrared spectra should be dominated by optically thick lines that only trace a molecular gas layer where disks are still optically thin in dust emission \citep[e.g. Figure 12 in][]{woitke18}, and this layer could also be radially located beyond an inner disk cavity \citep{anton16}.
The slab-model fit results are also known to depend on the spectral range of emission lines considered and to present an evident degeneracy, with higher temperature and lower column giving similar good fits to lower temperature and higher column, and typical uncertainties of 50--100~K in temperature and a factor of a few to ten in column density \citep{salyk11a,cn11}. Recent efforts have included radial gradients in gas temperature and density \citep{liu19} or full thermo-chemical modeling of inner disks \citep[e.g.][]{woitke18}, but have not been able to provide results for large disk samples yet. Therefore, despite their simplicity and degeneracy, for reference we report here the slab-model estimates from \citet{salyk11a}.

While fit results to individual disks might be questionable on the basis of the individual quality of data and fit, and overall scatters are still large, it is remarkable to find in Figure \ref{fig: S11_fits} that global trends appear where the column of water vapor is larger/smaller for smaller/larger R$_{\rm{dust}}$ and n$_{13-30}$ values (bear in mind that these results were obtained a decade ago and were agnostic of the analysis presented here). In the figure, we also include the radius of CO emission R$_{\rm{co}}$ as in Appendix \ref{app: IRindex}, in reference to recent work of spectrally-resolved near-infrared CO emission \citep{bp15,banz17}. The column of warm water vapor shows a trend with R$_{\rm{co}}$ too, consistent with a decrease in gas column density as the molecular gas recedes to larger radii in disks that have an inner dust cavity, supporting the interpretation proposed in \citet[][]{banz17}. 

While a lower \ce{H2O} column in disks with inner dust cavities was already clear from the analysis of \citet{salyk11a}, it is particularly interesting to note that disks that were found to have low water columns and yet no inner disk cavity are now shown to have large R$_{\rm{dust}}$. To better visualize this, in Figure \ref{fig: S11_fits} we use symbol sizes that are proportional to R$_{\rm{dust}}$. Again we remark that these estimates are degenerate, but the picture that emerges is consistent with the pebble drift interpretation discussed above in Section \ref{sec: disc_drift}. A clear venue for future work is to revisit water emission fits for a large sample of disks that span ranges of key disk parameters, and confirm whether the \ce{H2O}-R$_{\rm{dust}}$ relation corresponds to a column density or an elemental C/O relation, or else (Section \ref{sec: future}).


\begin{thebibliography}{}

\bibitem[Alcal{\'a} et al.(2017)]{alc17} Alcal{\'a}, J.~M., Manara, C.~F., Natta, A., et al.\ 2017, \aap, 600, A20

\bibitem[Alcal{\'a} et al.(2019)]{alc19} Alcal{\'a}, J.~M., Manara, C.~F., France, K., et al.\ 2019, \aap, 629, A108

\bibitem[Andrews, \& Williams(2005)]{aw05} Andrews, S.~M., \& Williams, J.~P.\ 2005, \apj, 631, 1134

\bibitem[Andrews et al.(2013)]{andr13} Andrews, S.~M., Rosenfeld, K.~A., Kraus, A.~L., et al.\ 2013, \apj, 771, 129

\bibitem[Andrews et al.(2016)]{andr16} Andrews, S.~M., Wilner, D.~J., Zhu, Z., et al.\ 2016, \apjl, 820, L40

\bibitem[Andrews et al.(2018a)]{andr18a} Andrews, S.~M., Terrell, M., Tripathi, A., et al.\ 2018a, \apj, 865, 157 

\bibitem[Andrews et al.(2018b)]{dsharp1} Andrews, S.~M., Huang, J., P{\'e}rez, L.~M., et al.\ 2018b, \apjl, 869, L41 

\bibitem[Andrews(2020)]{andr20} Andrews, S.~M.\ 2020, arXiv e-prints, arXiv:2001.05007

\bibitem[Antonellini et al.(2016)]{anton16} Antonellini, S., Kamp, I., Lahuis, F., et al.\ 2016, \aap, 585, A61

\bibitem[Antonellini et al.(2020)]{anton20} Antonellini, S., Banzatti, A., Kamp, I., et al.\ 2020, \aap, 637, A29

\bibitem[Arulanantham et al.(2018)]{aru18} Arulanantham, N., France, K., Hoadley, K., et al.\ 2018, \apj, 855, 98

\bibitem[Bae et al.(2018)]{bae18} Bae, J., Pinilla, P., \& Birnstiel, T.\ 2018, \apjl, 864, L26

\bibitem[Bailer-Jones et al.(2018)]{bj18} Bailer-Jones, C.~A.~L., Rybizki, J., Fouesneau, M., et al.\ 2018, \aj, 156, 58

\bibitem[Ballering \& Eisner(2019)]{ball19} Ballering, N.~P., \& Eisner, J.~A.\ 2019, \aj, 157, 144

\bibitem[Banzatti et al.(2012)]{banz12} Banzatti, A., Meyer, M.~R., Bruderer, S., et al.\ 2012, \apj, 745, 90

\bibitem[Banzatti(2013)]{banz13} Banzatti, A.\ 2013, Ph.D. Thesis, doi:10.3929/ethz-a-010093785

\bibitem[Banzatti et al.(2015)]{banz15} Banzatti, A., Pontoppidan, K.~M., Bruderer, S., et al.\ 2015, \apjl, 798, L16

\bibitem[Banzatti \& Pontoppidan(2015)]{bp15} Banzatti, A., \& Pontoppidan, K.~M.\ 2015, \apj, 809, 167

\bibitem[Banzatti et al.(2017)]{banz17} Banzatti, A., Pontoppidan, K.~M., Salyk, C., et al.\ 2017, \apj, 834, 152

\bibitem[Banzatti et al.(2018)]{banz18} Banzatti, A., Garufi, A., Kama, M., et al.\ 2018, \aap, 609, L2

\bibitem[Banzatti et al.(2019)]{banz19} Banzatti, A., Pascucci, I., Edwards, S., et al.\ 2019, \apj, 870, 76

\bibitem[Bate(2018)]{bate18} Bate, M.~R.\ 2018, \mnras, 475, 5618

\bibitem[Bergin \& van Dishoeck(2012)]{bvd12} Bergin, E.~A., \& van Dishoeck, E.~F.\ 2012, Philosophical Transactions of the Royal Society of London Series A, 370, 2778

\bibitem[Birnstiel et al.(2018)]{birn18} Birnstiel, T., Dullemond, C.~P., Zhu, Z., et al.\ 2018, \apjl, 869, L45

\bibitem[Blevins et al.(2016)]{blev16} Blevins, S.~M., Pontoppidan, K.~M., Banzatti, A., et al.\ 2016, \apj, 818, 22

\bibitem[Boogert et al.(2015)]{boog15} Boogert, A.~C.~A., Gerakines, P.~A., \& Whittet, D.~C.~B.\ 2015, \araa, 53, 541

\bibitem[Booth, \& Ilee(2019)]{booth19} Booth, R.~A., \& Ilee, J.~D.\ 2019, \mnras, 1425

\bibitem[Bosman et al.(2017)]{bosm17} Bosman, A.~D., Bruderer, S., \& van Dishoeck, E.~F.\ 2017, \aap, 601, A36

\bibitem[Bosman et al.(2018)]{bosm18} Bosman, A.~D., Tielens, A.~G.~G.~M., \& van Dishoeck, E.~F.\ 2018, \aap, 611, A80

\bibitem[Bosman et al.(2019)]{bosm19} Bosman, A.~D., Banzatti, A., Bruderer, S., et al.\ 2019, \aap, 631, A133

\bibitem[Bouvier et al.(2020)]{bouv20} Bouvier, J., Perraut, K., Le Bouquin, J.-B., et al.\ 2020, \aap, 636, A108

\bibitem[Brown et al.(2007)]{brow07} Brown, J.~M., Blake, G.~A., Dullemond, C.~P., et al.\ 2007, \apjl, 664, L107

\bibitem[Brown et al.(2013)]{brow13} Brown, J.~M., Pontoppidan, K.~M., van Dishoeck, E.~F., et al.\ 2013, \apj, 770, 94

\bibitem[Bruderer(2013)]{brud13} Bruderer, S.\ 2013, \aap, 559, A46

\bibitem[Carmona et al.(2017)]{carm17} Carmona, A., Thi, W.~F., Kamp, I., et al.\ 2017, \aap, 598, A118

\bibitem[Carr \& Najita(2011)]{cn11} Carr, J.~S., \& Najita, J.~R.\ 2011, \apj, 733, 102.

\bibitem[Ciesla, \& Cuzzi(2006)]{cc06} Ciesla, F.~J., \& Cuzzi, J.~N.\ 2006, \icarus, 181, 178

\bibitem[Cyr et al.(1998)]{cyr98} Cyr, K.~E., Sears, W.~D., \& Lunine, J.~I.\ 1998, \icarus, 135, 537

\bibitem[Cleeves et al.(2018)]{clee18} Cleeves, L.~I., {\"O}berg, K.~I., Wilner, D.~J., et al.\ 2018, \apj, 865, 155

\bibitem[Costigan et al.(2012)]{cos12} Costigan, G., Scholz, A., Stelzer, B., et al.\ 2012, \mnras, 427, 1344

\bibitem[Cox et al.(2013)]{cox13} Cox, A.~W., Grady, C.~A., Hammel, H.~B., et al.\ 2013, \apj, 762, 40

\bibitem[Cuzzi \& Zahnle(2004)]{cz04} Cuzzi, J.~N., \& Zahnle, K.~J.\ 2004, \apj, 614, 490

\bibitem[D'Alessio et al.(2006)]{dal06} D'Alessio, P., Calvet, N., Hartmann, L., et al.\ 2006, \apj, 638, 314

\bibitem[Du \& Bergin(2014)]{du14} Du, F. \& Bergin, E.~A.\ 2014, \apj, 792, 2

\bibitem[Dullemond et al.(2018)]{dull18} Dullemond, C.~P., Birnstiel, T., Huang, J., et al.\ 2018, \apjl, 869, L46

\bibitem[Facchini et al.(2019)]{facch19} Facchini, S., van Dishoeck, E.~F., Manara, C.~F., et al.\ 2019, \aap, 626, L2

\bibitem[Fang et al.(2018)]{fang18} Fang, M., Pascucci, I., Edwards, S., et al.\ 2018, \apj, 868, 28

\bibitem[Fedele et al.(2011)]{fed11} Fedele, D., Pascucci, I., Brittain, S., et al.\ 2011, \apj, 732, 106

\bibitem[Francis \& van der Marel(2020)]{fran20} Francis, L. \& van der Marel, N.\ 2020, \apj, 892, 111

\bibitem[Furlan et al.(2006)]{furl06} Furlan, E., Hartmann, L., Calvet, N., et al.\ 2006, \apjs, 165, 568

\bibitem[Furlan et al.(2009)]{furl09} Furlan, E., Watson, D.~M., McClure, M.~K., et al.\ 2009, \apj, 703, 1964

\bibitem[Gaia Collaboration et al.(2018)]{gaiaDR2} Gaia Collaboration, Brown, A.~G.~A., Vallenari, A., et al.\ 2018, \aap, 616, A1

\bibitem[Garufi et al.(2013)]{gar13} Garufi, A., Quanz, S.~P., Avenhaus, H., et al.\ 2013, \aap, 560, A105

\bibitem[Hendler et al.(2020)]{hend20} Hendler, N., Pascucci, I., Pinilla, P., et al.\ 2020, \apj, 895, 126

\bibitem[Herczeg \& Hillenbrand(2014)]{herc14} Herczeg, G.~J. \& Hillenbrand, L.~A.\ 2014, \apj, 786, 97

\bibitem[Honda et al.(2015)]{hond15} Honda, M., Maaskant, K., Okamoto, Y.~K., et al.\ 2015, \apj, 804, 143

\bibitem[Houck et al.(2004)]{IRS} Houck, J.~R., Roellig, T.~L., van Cleve, J., et al.\ 2004, \apjs, 154, 18

\bibitem[Huang et al.(2018)]{dsharp2} Huang, J., Andrews, S.~M., Dullemond, C.~P., et al.\ 2018, \apjl, 869, L42

\bibitem[Huang et al.(2020)]{huang20} Huang, J., Andrews, S.~M., Dullemond, C.~P., et al.\ 2020, \apj, 891, 48

\bibitem[Johansen \& Lambrechts(2017)]{joha17} Johansen, A., \& Lambrechts, M.\ 2017, Annual Review of Earth and Planetary Sciences, 45, 359

\bibitem[Kelly(2007)]{kelly07} Kelly, B.~C.\ 2007, \apj, 665, 1489

\bibitem[Krijt et al.(2018)]{krijt18} Krijt, S., Schwarz, K.~R., Bergin, E.~A., et al.\ 2018, \apj, 864, 78

\bibitem[Krijt et al.(2020)]{krijt20} Krijt, S., et al.\ 2020, submitted

\bibitem[Kurtovic et al.(2018)]{kurt18} Kurtovic, N.~T., P{\'e}rez, L.~M., Benisty, M., et al.\ 2018, \apj, 869, L44

\bibitem[Kurtovic et al.(2020)]{kurt20} Kurtovic, N.~T., Pinilla, P.~A., et al. submitted

\bibitem[Lambrechts et al.(2019)]{lamb19} Lambrechts, M., Morbidelli, A., Jacobson, S.~A., et al.\ 2019, \aap, 627, A83

\bibitem[Lebouteiller et al.(2015)]{cassis} Lebouteiller, V., Barry, D.~J., Goes, C., et al.\ 2015, \apjs, 218, 21

\bibitem[Liu et al.(2019)]{liu19} Liu, Y., Pascucci, I., \& Henning, T.\ 2019, \aap, 623, A106

\bibitem[Lodato et al.(2019)]{lod19} Lodato, G., Dipierro, G., Ragusa, E., et al.\ 2019, \mnras, 486, 453

\bibitem[Long et al.(2018)]{long18} Long, F., Pinilla, P., Herczeg, G.~J., et al.\ 2018, \apj, 869, 17 

\bibitem[Long et al.(2019)]{long19} Long, F., Herczeg, G.~J., Harsono, D., et al.\ 2019, \apj, 882, 49

\bibitem[Long et al.(2020)]{long20} Long, D.~E., Zhang, K., Teague, R., et al.\ 2020, \apjl, 895, L46

\bibitem[Loomis et al.(2017)]{loom17} Loomis, R.~A., {\"O}berg, K.~I., Andrews, S.~M., et al.\ 2017, \apj, 840, 23

\bibitem[Mac{\'\i}as et al.(2018)]{mac18} Mac{\'\i}as, E., Espaillat, C.~C., Ribas, {\'A}., et al.\ 2018, \apj, 865, 37

\bibitem[Manara et al.(2014)]{man14} Manara, C.~F., Testi, L., Natta, A., et al.\ 2014, \aap, 568, A18

\bibitem[Manara et al.(2016)]{man16} Manara, C.~F., Fedele, D., Herczeg, G.~J., et al.\ 2016, \aap, 585, A136

\bibitem[Manara et al.(2019)]{man19} Manara, C.~F., Tazzari, M., Long, F., et al.\ 2019, \aap, 628, A95

\bibitem[Mandell et al.(2012)]{mand12} Mandell, A.~M., Bast, J., van Dishoeck, E.~F., et al.\ 2012, \apj, 747, 92

\bibitem[Manset et al.(2009)]{mans09} Manset, N., Bastien, P., M{\'e}nard, F., et al.\ 2009, \aap, 499, 137

\bibitem[Morbidelli(2020)]{morbi20} Morbidelli, A.\ 2020, \aap, 638, A1

\bibitem[Mulders et al.(2015)]{muld15} Mulders, G.~D., Ciesla, F.~J., Min, M., et al.\ 2015, \apj, 807, 9

\bibitem[Mumma \& Charnley(2011)]{mum11} Mumma, M.~J., \& Charnley, S.~B.\ 2011, \araa, 49, 471

\bibitem[Najita et al.(2010)]{naj10} Najita, J.~R., Carr, J.~S., Strom, S.~E., et al.\ 2010, \apj, 712, 274

\bibitem[Najita et al.(2011)]{naj11} Najita, J.~R., {\'A}d{\'a}mkovics, M., \& Glassgold, A.~E.\ 2011, \apj, 743, 147

\bibitem[Najita et al.(2013)]{naj13} Najita, J.~R., Carr, J.~S., Pontoppidan, K.~M., et al.\ 2013, \apj, 766, 134

\bibitem[Najita et al.(2018)]{naj18} Najita, J.~R., Carr, J.~S., Salyk, C., et al.\ 2018, \apj, 862, 122

\bibitem[Notsu et al.(2017)]{notsu17} Notsu, S., Nomura, H., Ishimoto, D., et al.\ 2017, \apj, 836, 118

\bibitem[Pascucci et al.(2011)]{pasc11} Pascucci, I., Sterzik, M., Alexander, R.~D., et al.\ 2011, \apj, 736, 13

\bibitem[Pascucci et al.(2013)]{pasc13} Pascucci, I., Herczeg, G., Carr, J.~S., et al.\ 2013, \apj, 779, 178

\bibitem[Pascucci et al.(2016)]{pasc16} Pascucci, I., Testi, L., Herczeg, G.~J., et al.\ 2016, \apj, 831, 125

\bibitem[Pascucci et al.(2020)]{pasc20} Pascucci, I., Banzatti, A., Gorti, U., et al.\ 2020, \apj, in press

\bibitem[Pinilla et al.(2012)]{pin12} Pinilla, P., Birnstiel, T., Ricci, L., et al.\ 2012, \aap, 538, A114

\bibitem[Pinilla et al.(2018)]{pin18} Pinilla, P., Tazzari, M., Pascucci, I., et al.\ 2018, \apj, 859, 32

\bibitem[Pinilla et al.(2020)]{pin20} Pinilla, P., Pascucci, I., \& Marino, S.\ 2020, \aap, 635, A105

\bibitem[Pontoppidan et al.(2008)]{pont08} Pontoppidan, K.~M., Blake, G.~A., van Dishoeck, E.~F., et al.\ 2008, \apj, 684, 1323

\bibitem[Pontoppidan et al.(2010a)]{pont10a} Pontoppidan, K.~M., Salyk, C., Blake, G.~A., et al.\ 2010a, \apj, 720, 887 

\bibitem[Pontoppidan et al.(2010b)]{pont10b} Pontoppidan, K.~M., Salyk, C., Blake, G.~A., et al.\ 2010b, \apjl, 722, L173

\bibitem[Pontoppidan et al.(2011)]{pont11} Pontoppidan, K.~M., Blake, G.~A., \& Smette, A.\ 2011, \apj, 733, 84

\bibitem[Pontoppidan et al.(2014)]{pont14} Pontoppidan, K.~M., Salyk, C., Bergin, E.~A., et al.\ 2014, Protostars and Planets VI, 363

\bibitem[Pontoppidan et al.(2014)]{pont14fadi} Pontoppidan, K.~M. and Blevins, S.~M. 2014, Faraday Discussions, 168, 49

\bibitem[Pontoppidan et al.(2018)]{pont18} Pontoppidan, K.~M., Bergin, E.~A., Melnick, G., et al.\ 2018, arXiv e-prints, arXiv:1804.00743

\bibitem[Ricci et al.(2012)]{ric12} Ricci, L., Trotta, F., Testi, L., et al.\ 2012, \aap, 540, A6

\bibitem[Richards et al.(2018)]{rich18} Richards, S.~N., Moseley, S.~H., Stacey, G., et al.\ 2018, Journal of Astronomical Instrumentation, 7, 1840015

\bibitem[Rigliaco et al.(2015)]{rigl15} Rigliaco, E., Pascucci, I., Duchene, G., et al.\ 2015, \apj, 801, 31 

\bibitem[Rosotti et al.(2019a)]{ros19a} Rosotti, G.~P., Booth, R.~A., Tazzari, M., et al.\ 2019a, \mnras, 486, L63

\bibitem[Rosotti et al.(2019b)]{ros19b} Rosotti, G.~P., Tazzari, M., Booth, R.~A., et al.\ 2019b, \mnras, 486, 4829

\bibitem[Rothman et al.(2013)]{hitran} Rothman, L.~S., Gordon, I.~E., Babikov, Y., et al.\ 2013, \jqsrt, 130, 4

\bibitem[Salyk et al.(2009)]{salyk09} Salyk, C., Blake, G.~A., Boogert, A.~C.~A., et al.\ 2009, \apj, 699, 330

\bibitem[Salyk et al.(2011a)]{salyk11a} Salyk, C., Pontoppidan, K.~M., Blake, G.~A., et al.\ 2011a, \apj, 731, 130.

\bibitem[Salyk et al.(2011b)]{salyk11b} Salyk, C., Blake, G.~A., Boogert, A.~C.~A., et al.\ 2011b, \apj, 743, 112

\bibitem[Salyk et al.(2013)]{salyk13} Salyk, C., Herczeg, G.~J., Brown, J.~M., et al.\ 2013, \apj, 769, 21

\bibitem[Salyk et al.(2015)]{salyk15} Salyk, C., Lacy, J.~H., Richter, M.~J., et al.\ 2015, \apjl, 810, L24

\bibitem[Salyk et al.(2019)]{salyk19} Salyk, C., Lacy, J., Richter, M., et al.\ 2019, \apj, 874, 24

\bibitem[Schaefer et al.(2018)]{schae18} Schaefer, G.~H., Prato, L., \& Simon, M.\ 2018, \aj, 155, 109

\bibitem[Simon et al.(2016)]{sim16} Simon, M.~N., Pascucci, I., Edwards, S., et al.\ 2016, \apj, 831, 169

\bibitem[Tazzari et al.(2017)]{tazz17} Tazzari, M., Testi, L., Natta, A., et al.\ 2017, \aap, 606, A88

\bibitem[Testi et al.(2014)]{testi14} Testi, L., Birnstiel, T., Ricci, L., et al.\ 2014, Protostars and Planets VI, 339

\bibitem[Thalmann et al.(2016)]{talm16} Thalmann, C., Janson, M., Garufi, A., et al.\ 2016, \apjl, 828, L17

\bibitem[Tobin et al.(2015)]{tobin15} Tobin, J.~J., Looney, L.~W., Wilner, D.~J., et al.\ 2015, \apj, 805, 125

\bibitem[Tobin et al.(2020)]{tobin20} Tobin, J.~J., Sheehan, P.~D., Megeath, S.~T., et al.\ 2020, \apj, 890, 130

\bibitem[Trapman et al.(2019)]{trap19} Trapman, L., Facchini, S., Hogerheijde, M.~R., et al.\ 2019, \aap, 629, A79

\bibitem[Trapman et al.(2020)]{trap20} Trapman, L., Ansdell, M., Hogerheijde, M.~R., et al.\ 2020, \aap, 638, A38

\bibitem[Tripathi et al.(2017)]{trip17} Tripathi, A., Andrews, S.~M., Birnstiel, T., \& Wilner, D.~J.\ 2017, \apj, 845, 44 

\bibitem[van der Marel et al.(2016)]{vdmar16} van der Marel, N., van Dishoeck, E.~F., Bruderer, S., et al.\ 2016, \aap, 585, A58

\bibitem[van der Marel et al.(2018)]{vdmar18} van der Marel, N., Williams, J.~P., Ansdell, M., et al.\ 2018, \apj, 854, 177

\bibitem[Visser(2009)]{viss09} Visser, R.\ 2009, Ph.D. Thesis

\bibitem[Zhang et al.(2018)]{zhang18} Zhang, S., Zhu, Z., Huang, J., et al.\ 2018, \apjl, 869, L47

\bibitem[Zhao et al.(2020)]{zhao20} Zhao, B., Tomida, K., Hennebelle, P., et al.\ 2020, \ssr, 216, 43

\bibitem[Walsh et al.(2015)]{walsh15} Walsh, C., Nomura, H., \& van Dishoeck, E.\ 2015, \aap, 582, A88

\bibitem[White \& Ghez(2001)]{whi01} White, R.~J. \& Ghez, A.~M.\ 2001, \apj, 556, 265

\bibitem[Woitke et al.(2016)]{woitke16} Woitke, P., Min, M., Pinte, C., et al.\ 2016, \aap, 586, A103

\bibitem[Woitke et al.(2018)]{woitke18} Woitke, P., Min, M., Thi, W.-F., et al.\ 2018, \aap, 618, A57


\end{thebibliography}
\end{document}